\documentclass{aastex}
\usepackage{emulateapj5}

\shorttitle{Stellar Kinematics and Metallicities in Leo I}
\shortauthors{Koch et al.}

\slugcomment{Accepted for publication in the Astrophysical Journal}

\begin{document}

\title{Stellar kinematics and metallicities in the Leo I dwarf spheroidal 
galaxy -- wide field implications for galactic evolution} 
\author{Andreas Koch\altaffilmark{1,$^{\dagger}$}, Mark
I.~Wilkinson\altaffilmark{2}, Jan T.~Kleyna\altaffilmark{3}, 
Gerard F.~Gilmore\altaffilmark{2},  \\Eva K.~Grebel\altaffilmark{1},  
A.~Dougal Mackey\altaffilmark{2}, N.~Wyn Evans\altaffilmark{2}, 
and R.~F.~G. Wyse\altaffilmark{4}}
\email{akoch@astro.ucla.edu}
\altaffiltext{1}{Astronomical Insitute of the University of Basel, 
Department of Physics and Astronomy, 
Venusstr. 7, CH-4102 Binningen, Switzerland \\
$^{\dagger}$ Present address: UCLA, Department of Physics and Astronomy, 430 Portola Plaza, Los Angeles, CA 90095} 
\altaffiltext{2}{Institute of Astronomy, Cambridge University, 
Madingley Road, Cambridge CB3 0HA, UK}
\altaffiltext{3}{Institute for Astronomy, University of Hawaii, 2860 
Woodlawn Drive, Honolulu, HI 96822} 
\altaffiltext{4}{Department of Physics and Astronomy, Johns Hopkins
University, 3400 North Charles Street, Baltimore, MD 21218}

\begin{abstract}
We present low-resolution spectroscopy of 120 red giants in the
Galactic satellite dwarf spheroidal (dSph) Leo\,I, obtained with the
GeminiN-GMOS and Keck-DEIMOS spectrographs.  We find stars with
velocities consistent with  membership of Leo\,I out to 1.3 King
tidal radii. By measuring accurate radial velocities with a median 
measurement error of 4.6\,km\,s$^{-1}$   
 we find a mean
systemic velocity of 284.2\,km\,s$^{-1}$ with a global velocity
dispersion of 9.9\,km\,s$^{-1}$.  The dispersion profile is consistent
with being flat out to the last data point. We show that a
marginally-significant rise in the radial dispersion profile at a
radius of $3\arcmin$ is not associated with any real localized kinematical
substructure.  Given its large distance from the Galaxy, tides are not
likely to have affected the velocity dispersion, a statement we
support from a quantitative kinematical analysis, as we observationally
reject the occurrence of a significant apparent rotational
signal or an asymmetric velocity distribution. 
Mass determinations adopting both isotropic stellar velocity
dispersions and more general models yield a $M/L$ ratio of 24, which
is consistent with the presence of a significant dark halo with a mass
of about $3 \times 10^7\,M_{\odot}$, in which the luminous component
is embedded. This suggests that Leo\,I exhibits dark matter properties
similar to those of other dSphs in the Local Group. Our data allowed us also to
determine metallicities for 58 of the targets. We find a mildly metal poor 
mean of $-1.31$\,dex and a full spread covering 1\,dex. In contrast to 
the majority of dSphs, Leo\,I  appears to show no radial gradient in its
metallicities, which points to a negligible role of external influences 
in this galaxy's evolution.
\end{abstract}

\section{Introduction}
Leo\,I is the most remote dwarf spheroidal (dSph) galaxy generally
believed to be associated with the Milky Way (MW) subsystem in the
Local Group (see Table 1 in Grebel, Gallagher, \& Harbeck 2003).  Due
to its remarkably high radial velocity of about 287 km~s$^{-1}$
(Zaritsky et al.\ 1989; Mateo et al.\ 1998) and its current large distance of
$\sim 254$\,kpc (Bellazzini et al.\ 2004), Leo\,I is a crucial tracer
for determining the total mass of the Galaxy (e.g., Kochanek 1996).
The question of whether Leo I is actually bound to the Galaxy remains
open (e.g., Byrd et al.\ 1994).  A quantitative assessment of the
possible role of Galactic tides in the internal (dynamical,
star-formation history, etc.) evolution of the Galactic satellite dSphs
is of considerable importance, for its implications for the evolution of
small galaxies, and for the distribution of dark matter on small
scales. In particular, Leo\,I is of great interest here, given its very large
Galactocentric distance, as it is the least-likely dSph to be
affected by tides
(but see also the recent work of Sohn et al. 2006, hereinafter S06). 
Similarities in its dynamics with those of more
nearby dwarfs would strengthen our confidence that Galactic tides are
not a major effect in determinations of the dark
matter distribution both of Leo\,I itself and the other dSphs 

The multiplexing capabilities of present-day multi-object
spectrographs on 4--10\,m class telescopes have opened a window to
study dSph galaxies in unprecedented detail by enabling one to gather
data sets for a large number of individual stars. As a result, the
systemic velocities of the majority of the Local Group (LG) dSphs are
known (tabulated in the reviews by Grebel 1997; Mateo 1998; van den
Bergh 1999; and more recent measurements as quoted in, e.g., Evans et
al.\ 2000, 2003).  Moreover, high-precision radial velocity data for many
individual stars in all the known dSphs are being obtained.  These
serve as the kinematical input for detailed dynamical studies (Mateo
et al.\ 1998; C{\^o}t{\'e} et al.\ 1999; Kleyna et al. 2001, 2003,
2004; \L okas et al. 2002; Wilkinson et al.\ 2004; Tolstoy et al.\
2004; Chapman et al.\ 2005; Mu\~noz et al.\ 2005, 2006a; Walker et al.\
2006a, 2006b; Wilkinson et al.\ 2006a,b; and references therein),
which are determining the spatial distribution of dark matter on the
smallest available scales.

DSphs are the least massive and least luminous galaxies known to
exist.  They are characterized by absolute $V$-band luminosities
$M_V\gtrsim-$14\,mag, low surface brightnesses of
$\mu_V\gtrsim$22\,mag\,arcsec$^{-2}$ and \ion{H}{1} masses of less
than $10^5$ M$_{\odot}$ (Grebel et al.\ 2003 and references therein).
They have high central stellar radial velocity dispersions, typically
of order $10$\,km\,s$^{-1}$ (Aaronson 1983; Mateo 1998), which,
together with their globular-cluster-like low luminosities (Gallagher
\& Wyse 1994; Mateo 1998) and their large core size -- hundreds of parsecs
rather than the few parsecs of star clusters -- demonstrate that these
galaxies are dominated by dark matter at all radii (Mateo et al. 1997;
Wilkinson et al. 2004, 2006a). This deduction presumes the dSph are in
a state of dynamical equilibrium, a presumption supported by direct
evidence that the dSph are not highly elongated down our lines of
sight (e.g., Odenkirchen et al. 2001; Mackey \& Gilmore 2003; Klessen
et al. 2003). This asumption merits careful test, however, especially
in the outer-most parts of the dSph, where Galactic tides might become
significant. 

The inferred high mass-to-light ratios ($M/L$), obtained under the
assumption of dynamical equilibrium, are as large as
500\,($M/L$)$_{\odot}$ or more, making dSphs the smallest observable
objects available to study the properties of dark matter halos.
Available measurements suggest that the dark matter in dSphs has some
common properties, including a minimum halo mass scale of $\sim
4\times 10^7$ M$_{\odot}$, a small range in central (dark) mass
densities, and a minimum length (core radius) scale of order
100\,pc (Wilkinson et al.\ 2006a). If established, these are the
first characteristic properties of dark matter.

Cold dark matter (CDM) simulations of galaxy formation predict the
existence of a large number of  sub-haloes surrounding larger
galaxies, such as the Milky Way or M\,31 (e.g., Moore et al.\ 1999).
The correctness of this prediction remains a subject of debate:
the number of observed dSphs is much less, though several have been
discovered very recently, indicating that the known sample is incomplete
(Belokurov 2006a, 2006b; Martin et al. 2006; Zucker et al. 2006a, 2006b, 2006c). The correspondence 
between prediction and
observation more importantly requires masses, so that dynamical
studies of the dSphs are critical. 

The spatial distribution of dark matter on small scales is an
additional test of the nature of dark matter. Numerical simulations of
dark matter halos in which it is assumed that there is no physical process
except gravity acting remain limited in spatial resolution, but in
general are characterized by a central density cusp $\rho(r)\propto
r^{-a}$ with $a$=1--1.5 (e.g., Navarro, Frenk \& White 1995, hereafter
NFW; Ghighna et al.\ 2000).  It is far from clear that these
expectations are consistent with observations of galaxies
significantly more massive than the dSphs (de Blok et al.\ 2001;
Spekkens et al.\ 2005), which are better accounted for by flat density
cores.  The available data on dSph kinematics appear to be consistent
with cored halos as well (\L okas 2002; Read \& Gilmore 2005;
Wilkinson et al.\ 2006a).  In the Ursa Minor dSph a central,
constant-density core has been detected (Kleyna et al.\ 2003), and
also in the Fornax dSph a core may be more likely than a cusped
distribution, although these inferences are still subject to large
uncertainties (Strigari et al.\ 2006; Goerdt et al.\ 2006).  It has
been suggested that the cusp/core distinction could be the first
indications of the properties of the physical particles which make up
CDM (Read \& Gilmore 2005), though it is essential that one
allows for dynamical evolution associated with the astrophysical
evolution of the dSph progenitors (Grebel et al.\ 2003).

According to a second 
scenario, the large observed velocity
dispersions may be explained if dSphs are unbound tidal remnants that
underwent tidal disruption while orbiting the Milky Way (Kuhn 1993;
Kroupa 1997; Klessen \& Kroupa 1998; Klessen \& Zhao 2002; Fleck \&
Kuhn 2003; Majewski et al.\ 2005), and all happen to be aligned to
mimic a bound system.
Predictions following from this scenario include a significant depth extent 
along the line of sight (Klessen \& Kroupa 1998;
Klessen \& Zhao 2002).  However, in two of the closest Milky Way
companions, the Draco dSph (Klessen et al.\ 2003) and the Fornax dwarf
(Mackey \& Gilmore 2003) direct evidence against such effects is
available.  Other arguments against dSphs being unbound tidal remnants
without dark matter include that they all experienced extended star
formation histories with considerable chemical  enrichment and that they follow
a metallicity-luminosity relation (see Grebel et al.\ 2003 for
details).
In the light of its large Galactocentric distance, Leo\,I is well
suited as a test of the possible importance of tidal perturbation.

The subject of tides in the context of dSph galaxies suffers from a
very distorted literature. In large part this follows from an
historical labelling issue. Specifically, 
the projected surface brightness distributions of
dSph galaxies are nowadays preferentialy fit with King models.
 A King model has two named linear
scales: an inner scale, frequently called a `core radius' and an outer
truncation radius, called a `tidal radius'. While these labels have
some relevance in their historical application to models of globular
clusters of stars truncated by Galactic tides, they have no general
physical interpretation in a different context. In general, dSph
stellar populations are not single-component, and there is no
astrophysical reason to assume ab initio that the distribution
function is simply isothermal. Some form of numerically tractable
functional fit to a projected surface brightness profile is required
in the dynamical analysis, but a more sophisticated analysis is
required than fitting functions to surface brightness profiles before
the possible importance of tides can be investigated. 
It is sometimes
suggested that {\em high-frequency} distorted photometric structure is
indicative of tidal damage, but no available numerical simulations
support this view: rather, if tides are important, they produce smooth
systematic distortions in the outer parts of a system. 
One test,
applicable in some circumstances, is detection of apparent rotation or 
spatially asymmetric velocity distributions 
associated with such a distortion, as claimed to be present in Leo\,I 
by S06. 
We will consider that in our analysis of Leo\,I.
Some of these relevant issues are discussed further by Read \& Gilmore (2005),
and by Read et al. (2006a, 2006b). 

Based on its projected position on the sky, Leo\,I was suggested to be
part of a Galactic Fornax-Leo-Sculptor Stream (Lynden-Bell 1982;
Majewski 1994), i.e., that it lies along a Galactic polar great plane
together with these other dSph galaxies.  However, despite the lack of
proper motion and orbital measurements for Leo\,I, the radial velocity
and kinematical data to hand apparently rule out a physical
association of Leo\,I with such a physical group (Lynden-Bell \&
Lynden-Bell 1995; Piatek et al. 2002, 2006). 
The remote Leo\,I dSph holds a special place in the kinematic studies
of MW satellites (Mateo et al. 1998, hereinafter M98).  From detailed
kinematical modeling of the velocity dispersion profile of 33 stars
within the core region of Leo\,I, M98 concluded that this dSph
contains a significant amount of dark matter.

Apart from its stellar kinematics, Leo\,I is interesting in terms of
its star formation history (SFH; see also Bosler et al. 2006). The dSphs of the Local Group exhibit
a variety of SFHs, where no two dwarfs are alike (Grebel 1997; Mateo
1998). In spite of this diversity, all nearby galaxies studied in
sufficient detail have been shown to share a common epoch of ancient
star formation, although the fraction of old populations with ages $>
10$ Gyr varies from galaxy to galaxy (Grebel 2000; Grebel \& Gallagher
2004).  Indeed Leo\,I has a prominent old stellar population (Held et
al.\ 2000), but more striking is the prevalence of intermediate-age
populations -- about 87\% of its stars formed between 1 and 7\,Gyr ago
(Gallart et al.\ 1999a).  The details of the SFH of Leo\,I were derived
photometrically, but particularly in galaxies with mixed populations
the age-metallicity degeneracy cannot be broken without spectroscopy.
Our current inferences of the chemical evolutionary histories of
nearby dwarf galaxies are still mainly based on photometry, while
detailed and numerous spectroscopy of as many stars as possible is
invaluable for helping to securely constrain the SFHs or to quantify
age-metallicity relations.  It is also essential to study the chemical
evolution of dSphs in order to understand their complex SFHs in terms
of regulating processes such as the inflow or accretion as well as the
outflow of gas (Carigi et al.\ 2002; Dong et al.\ 2003; Lanfranchi \&
Matteucci 2004; Hensler, Theis, \& Gallagher 2004; Robertson et al.\
2005; Font et al.\ 2006).  We are currently conducting a program to
measure the kinematics and chemical abundances in several Galactic
dSphs and to correlate these with their evolutionary histories (e.g.,
Wilkinson et al.\ 2006b, Koch et al.\ 2006a).

Here we present a kinematic and abundance study of the distant
Galactic dSph Leo\,I. In addition to a detailed kinematic analysis,
appropriate to the size of the available dataset, we have measured the
average metallicity and the spread and shape of Leo\,I's metallicity
distribution function (MDF), which provide valuable insights on any
possible environmental dependence of the processes governing the
evolution of dSphs.
This paper is organized as follows: \textsection 2 describes our data 
acquisition and reduction and the determination of stellar velocities.  
In \textsection 3 we derive Leo\,I's global 
velocity distribution from which we investigate potential  
velocity gradients in \textsection 4. Section 5 is then dedicated to 
the analysis of Leo\,I's radial velocity dispersion profile 
and the influence
of binaries on this profile. This radial profile allowed us to 
obtain mass and density estimates of this galaxy in \textsection 6. 
The role of an anisotropic velocity distribution on dispersion and mass estimates 
is addressed in \textsection 7.  
In \textsection 8 we derive metallicities for our targets and discuss 
the implications for Leo\,I's chemical evolution, and \textsection 9 
finally summarizes our results. 

\section{Data}
\subsection{Target selection}
Targets were selected from photometry obtained within the framework of
the Cambridge Astronomical Survey Unit (CASU) at the 2.5\,m Isaac
Newton Telescope (INT) on La Palma, Spain\footnote{see
\url{http://www.ast.cam.ac.uk/$\sim$wfcsur/}.}.  From this, we
selected red giants so as to cover magnitudes from the tip of the red
giant branch (RGB) at $I\sim18$\,mag (Bellazzini et al. 2004) down to
$\sim$1.6\,mag below the RGB tip, going as faint as $I\lesssim19.6$.
At our faintest observed magnitudes, our spectra reach signal-to-noise
(S/N) ratios of $\sim$5, which is sufficient to enable
accurate radial velocity measurements at our low resolution.

\subsection{Observations}
One part of our observations was carried out in queue mode with
the Gemini Multiobject Spectrograph GMOS at the 8-meter Gemini North
telescope, Hawaii, over four nights between December 2004 and January
2005.  These data were obtained under photometric conditions and a
clear sky.  The camera consists of three adjacent 2048$\times$4068
CCDs, which are separated by gaps with a width of $\sim$37\,pixel
each.  We used 0$\farcs$5 wide multiobject slits and the R400+G5305
grating set in two modes, with central wavelengths of 8550\,\AA\ and
8600\,\AA, providing a spectral coverage of 1800\,\AA.  This dithering
enabled us to efficiently interpolate the spectra across the CCDs'
gaps during the later coaddition process. The CCDs were binned by
2$\times$4 during readout and the resulting spectra have a nominal
resolution of 2200.  In total, we targeted 68 red giants spread over
three different fields in the eastern half of the galaxy (see Fig.~1,
Table~1).  Each of the three fields was exposed for three hours in
total, where we split up the exposures into $3\times 1800$\,s per
setup (i.e., each of the two wavelength modes) to facilitate sky
subtraction and cosmic ray removal. Due to bad weather, the central
pointing (Field 3, see Table~1) was exposed for only 3600\,s in total.

The other part of our data was obtained with the DEep Imaging
Multi-Object Spectrograph (DEIMOS) at the 10-meter Keck\,II
telescope. 
Leo I was observed on the DEIMOS multislit spectrograph on the Keck II
telescope on February 6 to February 8, 2005. Conditions were mediocre,
with high humidity and poor seeing, but approximately six hours of
Leo I observations were obtained, consisting of five slit masks concentrated
on the dSph's periphery. We used the 1200 line/mm grating. The slits 
were $0.7^{\prime\prime}$ in size, and
our resolution was about 0.3 \AA\ per pixel.

\subsection{Data Reduction}
\subsubsection{GMOS data}
The GMOS data were reduced using the standard GMOS reduction package
in IRAF, starting with the bias frames taken as part of the day-time
calibration process. Spectroscopic flat-field exposures were obtained
adjacent to each set of science exposures, from which we could model
and remove the spectral signature and spatial profile of the
illumination pattern. Through dividing by the remaining normalized
flat-field frame, we ensured a complete removal of any residual
sensitivity variation.
Since the CuAr wavelength calibration frames were taken up to four
days before and/or after the science spectra, during which time the
telescope flexure may significantly change, we had to rely on the
skylines of each individual science exposure to obtain an accurate
wavelength solution.  Hence, we chose a number of strong and isolated,
unblended night-sky OH-emission lines around the CaT region from the
atlas of Osterbrock et al. (1996, 1997). Fitting the respective lines
with a low- order polynomial yielded a RMS uncertainty in the
wavelength calibration of the order of 0.05\,\AA, corresponding to
$\sim 2$\,km\,s$^{-1}$ at our region of interest\footnote{Using the
entire spectral region at hand, covering ca. 1800\,\AA, yielded highly
inaccurate results in the wavelength solution, as too large a range of
distortions was being fitted with too few lines, resulting in offsets
between individual exposures of the order of $\sim
50-100$\,km\,s$^{-1}$. However, when concentrating on the narrow
window ($\sim$500\AA) around the CaT, in which we are primarily
interested, we circumvent any distortions towards the edges of the
spectra and could thus obtain a highly improved accuracy.}.
A crucial step in the analysis of CaT absorption features consists of
the accurate subtraction of the prominent skylines in the
near-infrared spectral region.  While the centers of the first two of
the Ca triplet lines are unaffected by adjacent skylines at Leo\,I's
systemic velocity, the third Ca absorption line coincides with the
dominant OH-band at $\sim$8670\,\AA.  These undesired contaminants are
efficiently removed by the software by fitting a low order polynomial
to the background and subtracting this fit column by column from the
science spectra.  Moreover, the accuracy of our sky subtracted spectra
(after coaddition) was improved by the fact that we obtained spectra
with the slit centred both at the centre and offset towards the right
and left hand side of the stellar seeing disk, yielding a better
coverage of the average sky background.  Finally, the extracted
spectra were shifted towards the local standard of rest barycenter and
median-combined using a standard $\sigma$-clipping algorithm.  
The average
S/N ratio achieved throughout our reductions is  $\sim$28, with a
minimum of 5.3 at I\,=\,19.6.
\subsubsection{DEIMOS data}

The DEIMOS data were reduced using the DEEP\footnote{see 
\url{http://astro.berkeley.edu/~cooper/deep/spec2d/}} pipeline, 
developed  by the DEEP2
project at the University of California-Berkeley
and based on the IDL spectral reduction 
pipeline of the Sloan Digital Sky Survey (Abazajian et al. 2004).   
The DEEP pipeline first identifies 
spectra using tungsten lamp flatfield exposures, and then finds a two 
dimensional (pixel-by-pixel)  wavelength solution for each spectrum, 
assisted by a DEIMOS-specific optical model.  Typically, the solution 
is accurate to 0.005\AA.  
All calibration images were taken at the start of each night; this is 
possible because the spectrograph optical system is 
stabilized  using an active flexure  compensation system.  Sky 
subtraction is performed  using a $b$-spline sky model.  
In the last stage, the sky-subtracted  two-dimensional spectra are 
extracted into one-dimensional flux and variance spectra   
in the form of FITS tables. The variance  spectrum is finally used to compute 
the spectral S/N ratio. The S/N achieved throughout our processing  
reaches $\sim$45 for the brighest targets and is as low as 5 for the faintest 
stars at I\,=\,21.2.

Generally, the DEEP pipeline worked as expected. For certain 
exposures, however, the slit identification and tracing in the flats failed, 
and we found that the sensitivity parameter controlling slit detection 
had to be adjusted.

\subsection{Radial Velocities}

Radial velocities were derived from our final reduced set of spectra
by cross-correlating the three strong Ca lines at $\lambda\lambda
8498,8542,8662$ against a synthetic template spectrum of the CaT
region using IRAF's cross-correlation package {\sc fxcor}.  The
template was synthesized using representative equivalent widths of the
CaT in red giants.  The final velocity difference between object
spectrum and the rest-frame template was then determined from a
parabolic fit to the strongest correlation peak.

\subsection{Velocity errors}
The formal velocity errors returned from the cross-correlation in IRAF
have a median value of 1.7\,km\,s$^{-1}$. However, we have found in
past studies that the {\sc fxcor} estimates are generally too
optimistic (see e.g., M98; Kleyna et al. 2002).  Since we have only single epoch data at
hand, lacking the possibility to compare the velocity stability over a
longer time period, we cannot estimate the true uncertainties through
the discrepancies of measurements taken at different times (Vogt el
al. 1995, M98). For the GMOS data, we rather follow the
prescription of Kleyna et al. (2002) and divide the data of the
brightest stars (with $I\le19$\,mag) into two final sets of spectra,
the first comprising those individual spectra with a central
wavelength of 8550\,\AA~and the second being those centered at
8600\,\AA. Each of these sets was separately cross-correlated against
the template. We use the discrepancy between the two resultant
velocities as an estimate of the true measurement error.  Hence, we
rescaled the Tonry-Davis $R$-based errors (Tonry \& Davis 1979) 
returned by {\sc fxcor} to
obtain the expected reduced $\chi^2$ discrepancy of unity. We found
that we had to apply a scale factor of 2.7 to yield an accurate
estimate of our velocity errors.
In order to check whether our wavelength calibration, which is based
on only a narrow spectral window around the CaT has introduced any
bias, we also determined radial velocities by cross-correlating each
of the three Ca lines separately against the same template. In this
case, the mean deviation of the resultant velocities did not exceed
1.8\,km\,s$^{-1}$.  These uncertainties were added in quadrature to
the formal measurement errors (including the above Tonry-Davis $R$
scaling) to yield the final velocity error, which we show in Table~2.
The final GMOS data set contained 68 stars with a median velocity error 
of 5.0\,km\,s$^{-1}$.

For the DEIMOS data, the nominal IRAF velocity errors were also
rescaled by a common factor.  The rescaling was performed by computing
separate velocities for lines 1+2 and line 3 of the CaT, and then
requiring the difference between the two velocities divided by the
quadrature sum of the rescaled nominal errors to obey the expected
$\chi^2$ statistics.  The final DEIMOS member list includes 27 member
velocities with per pixel $S/N>10$, and a median velocity error of 2.7
$\rm km\,s^{-1}$.

\section{Radial Velocity Distributions}
To determine an overall mean velocity and a global dispersion for
Leo\,I, a prerequisite for an assessment of the targets' membership, we used
the maximization method described  in Kleyna et al. (2002) and
Walker et al. (2006a).  In this approach, the probability that an observed
distribution of velocities $v_i$ and corresponding measurement errors
$\sigma_i$ is drawn from a Gaussian distribution with mean
$\left<v\right>$ and dispersion $\sigma$ is given by
\begin{equation}
p\,(\{ v_i,\sigma_i \} | (\left<v\right>,\sigma)\,)\,\propto\,  \\
\Pi_{i=1}^N\,\frac{1}{\sqrt{2\pi(\sigma^2\,+\,\sigma_i^2)}}\,\exp\left\{
-\frac{(v_i\,-\,\left<v\right>)^2}{2\,(\sigma^2\,+\,\sigma_i^2)}\right\}
\end{equation}
It is mathematically more convenient to calculate the natural
logarithm of this expression, and maximizing $\ln p$ is equivalent to
maximizing $p$ itself.  Rejecting stars deviating more by than
3$\sigma$ as likely non-members, we iterated until convergence (requiring that the 
solution from subsequent iterations changed by no more than 0.005\,km\,s$^{-1}$).  The
respective uncertainties in the mean and dispersion were then
calculated from the covariance matrix (Walker et al. 2006a).  This
procedure yielded a mean systemic velocity of Leo I of (284.2$\pm$
1.0)\,km\,s$^{-1}$ and a dispersion of 9.9\,$\pm$1.5\,km\,s$^{-1}$
from the combined GMOS plus DEIMOS sample, (284.6$\pm$1.5,
9.9$\pm$2.1)\,km\,s$^{-1}$ for the GMOS data alone and (283.8$\pm$1.5,
9.9$\pm$2.1)\,km\,s$^{-1}$ as derived from the DEIMOS targets only.
The mean velocities and global dispersions are tabulated in Table~3
separately for the different fields and Fig.~2 shows the resulting
histograms.

This compares to a value of $287.0\,\pm\,1.9$\,km\,s$^{-1}$ as found
by M98 and their respective dispersion of $8.8\,\pm
1.3$\,km\,s$^{-1}$ for stars in the innermost $3\farcm 5$.

If we take the $3\sigma$-cut as a membership criterion, 5 of the 68
GMOS targets and 5 of the red giants from the DEIMOS sample have to be
rejected as foreground contamination, where 4 out of these 5 stars in
each sample have velocities below 200\,km\,s$^{-1}$.
In order to generate the final sample for determining the dispersion
profile of Leo\,I we combined both of our GMOS and DEIMOS sets by
shifting them to a common median velocity.  This procedure is
justified by the lack of any significant radial gradient in the
velocities from DEIMOS, which is found to be
($-0.04\pm0.18)$\,\,km\,s$^{-1}$ arcmin$^{-1}$ so that such a shift
does not introduce any artificial velocity gradient or a falsification
of the dispersion profiles.
At the end of this procedure, we found that our median velocity 
from the GMOS plus DEIMOS set 
differed from that of M98 by $-$2.8 $\rm km\,s^{-1}$, a
difference significant only at the $p=0.31$ level (note that 
our sample has no stars in common with the M98 sample, which 
would allow for a direct comparison of the velocities).  
These values are finally listed in Table~2.  
\section{Kinematic tests for tidal damage and apparent rotation}

Those dSph galaxies studied so far show no significant contribution
from angular momentum to the systems' dynamical structure: dSphs obtain
their size and shape from anisotropic random pressure.  This is,
incidentally, an argument against an evolutionary transformation from
dwarf irregular to dSph galaxies via ram pressure stripping (e.g., 
van den Bergh 1999; Grebel et al. 2003).
More fundamentally, apparent rotation (generally visible in projection) is 
a characteristic signature of tidal disturbance. In this context, 
apparent rotation of the stellar component of a dSph is a test of
possible tidal `heating' of the stellar orbits in the course of the
galaxy's orbit within the external tidal field of the Milky Way.
N-body simulations, e.g., of Oh et al. (1995) have shown that the
velocity dispersion of a tidally-limited system is sustained at the
actual virial equillibrium value.  On the other hand, these
simulations suggest that any strong dynamical effects on a dSph from
the tidal field of the Galaxy can invoke streaming motions in the
outskirts of the dwarf, which will lead to a systematic change in the
mean velocity along its major axis, thus mimicking rotation
(Piatek \& Pryor 1995; Oh et al. 1995; Johnston et al. 1995; M98). A
more detailed analysis of tidal effects on dSph galaxies is provided
by Read et al. (2006b).

It has frequently been suggested in the literature that the detection
of member stars well beyond the formal tidal-limit radius of dSphs,
canonically defined via a single-component King model representation of
the observed surface brightness profiles, is evidence for physical
tides. A large number of dSphs have been claimed to show hints of
extratidal stars and thus tidal disruption (Irwin \& Hatzidimitriou
1995; Mart\'inez-Delgado et al. 2001; Palma et al. 2003; Mu\~noz et
al. 2005, 2006a, 2006b). Evidence for this has also been reported 
for Leo\,I (S06).  
What such studies show, in fact, is that the
presumed parametric fit to the measured surface brightness (and/or
direct star counts) is inappropriate at large radii. 
Given that there is no astrophysical
basis to the application of a King model to a galaxy, 
an astrophysical interpretation of the corresponding labels
as physical processes is fraught with peril.  
Evidence that a specific dSph is being
affected by tides sufficiently strongly that its kinematics are
affected can be provided most efficiently from a kinematic analysis. A test for
apparent outer rotation is such a test.

With its present-day Galactocentric distance of (254$\pm$19)\,kpc and
its high systemic velocity, Leo\,I is not expected to be affected by
tides. M98 concluded from their kinematical data that the core of this
galaxy has not been significantly heated by tides.  
S06 report on a large population of photometrical and radial
velocity member red giant stars out to nearly two nominal tidal radii,
which they interpret as evidence of tidal disruption. 
We note that also our sample comprises nine red giants outside $r_{tid}$,
whose velocities suggest membership. 
We also emphasise that these
statements tell us that the outer parts of Leo\,I are not well-fit by
a single King model. Whether they tell us anything about the dynamics of Leo\,I 
remains questionable for the moment and has to await further dynamical tests.

In order to assess whether there is any significant indication of
rotation in the Leo I velocities, which may help to elucidate the role
of tides, we pursued two tests. 
First, we show in Fig.~3 the spatial distribution of our measured radial 
velocities (including M98's 33 data points) along their 
projected major axis radial distance. In addition, we calculated the 
mean velocity as a function of radius, which is shown in the lower panel 
of Fig.~3. It is worth noticing that  this mean velocity remains 
fairly constant across the range covered by our data and the deviations 
from the global mean of our sample are well within the errorbars.  
Moreover, we do not detect any significant sign of a velocity gradient 
in the inner regions as suggested by S06. 
If Leo\,I has been significantly affected  by Galactic tides, one would 
expect an asymmetric excess of velocity outliers, with stars of higher 
or lower velocity typically lying on opposite sides of the galaxy 
due to the presence of possible  leading and trailing 
tidal tails. 
In this vein, S06 report on the 
presence of a number of stars with higher velocities than the average 
in the outer regions to the western half 
of Leo\,I. The tidal model then predicts a comparable presence of low 
velocity stars towards the eastern extension. Although our sample 
contains one red giant at a large (western) distance with a radial velocity 
that is higher by  $3.8\,\sigma$ than the sample mean, there is no apparent 
tendency of a systematically lower mean velocity of the targets at large distances towards 
the east.

While we find a mean radial velocity of 294.3$\pm$2.9\,km\,s$^{-1}$ 
for the stars outside the formal tidal radius on the eastern side, 
the value for the stars towards the western half including (rejecting) 
the high velocity outlier amounts to 289.5$\pm$4.0 (285.7$\pm$5.5) km\,s$^{-1}$. 
From this point of view our data do not show any hint of significant rotation, 
nor an asymmetry of the velocities on either side along the galaxy's major axis. 

Secondly, we 
calculated the difference in mean velocity on either side
of bisecting lines passing through each of the targets, assuming that
each such line is a potential rotation axis.  The axis yielding the
maximum mean velocity difference is then taken as the ``rotation
axis'', with the associated velocity difference being an estimate of
the amplitude of the rotation.
Fig.~4 shows the resulting velocity differences for our dataset.  Due
to the confinement of our GMOS fields to the eastern half of Leo\,I,
we cannot perform any test for rotation in this subsample
alone. However, based on only the 56 candidate members from DEIMOS, we
find a maximum velocity difference of 1.2\,km\,s$^{-1}$. Nevertheless, the
spatial distribution of these stars does not allow one to
unambiguously determine a potential rotation axis (see top panel of
Fig.~4).
In the case of the combined GMOS and DEIMOS sample, the amplitude of
the potential rotation reaches a value of 5.2\,km\,s$^{-1}$ (middle
panel of Fig.~4). When also the 33 stars from M98 are added, the mean
velocity difference changes to 3.0\,km\,s$^{-1}$ (bottom panel of
Fig.~4).  It is worth noticing that in both of the latter cases the
axis of the maximum rotation signal occurs at a position angle
(78$\degr$, 80$\degr$) which coincides with Leo\,I's
position angle of $(79 \pm 3)\degr$ (Irwin \& Hatzidimitriou 1995).

In order to test the significance of these results, we constructed
10$^4$ random samples of radial velocities at the observed spatial
positions.  The velocities were taken from a normal distribution with
the same mean and standard deviation as in our observations,
additionally allowing for a variation due to the measurement
uncertainty.
By means of this Monte Carlo simulation we find that the
significance of the rotation in the DEIMOS sample is consistent with
zero, whereas the amplitudes of the combined GMOS plus DEIMOS data
sets and the full data including M98's stars amount to a significance
of 86.7\% (1.5\,$\sigma$) and 20\% (0.3\,$\sigma$), respectively. Thus
the rotational signature we detected appears not to be statistically
significant in either case. Consequently, we have no evidence of any
kinematical effect due to Galactic tides operating on Leo\,I.

Another cause of apparent rotation is relative transverse motion.
Unfortunately, there is no estimate of Leo I's proper motion available
in the literature at present.  Hence, we cannot rule out the
possibility that the amount of observed rotation is entirely due to
the relative motion between the Sun and the dSph, $v_{rel}(l,b)$, in
the Galactocentric rest frame.  A non-rotating object will appear to
rotate as seen in the heliocentric rest frame, provided there is a
significant gradient in this relative motion across the dSph (see
Walker et al. 2006a).  We can attempt the inverse calculation: we
tested whether one can constrain Leo\,I's true proper motion by
demanding that this galaxy does not rotate but the entire amplitude of
the observed velocity gradient is caused by such projection effects.
For this purpose, we calculated the Galactocentric rest frame velocity
of each star through the formalism provided in Piatek et al. (2002) and
Walker et al. (2006a). This was done for a wide grid of assumed proper
motions, corresponding to transverse velocities from $-500$ to
+500\,km\,s$^{-1}$ in both $\mu_l$ and $\mu_b$. Regrettably, the full
range of input ($\mu_l,\,\mu_b$) was able to reproduce the observed
apparent rotation amplitude adequately, so no deductions are possible.
We then re-ran the same formalism by adopting a toy distance of
50\,kpc to Leo\,I and secondly assumed its line-of-sight velocity as
zero. In both cases the parameter space of possible proper motions
increased even more. In consequence, we can neither unequivocally
exclude the possibility that Leo\,I's observed, though insignificant,
velocity gradient is purely caused by the aforementioned projection
effects or due to its large distance or systemic
velocity, nor can we conclusively restrict its proper motion to 
a reasonable estimate.
\section{Velocity Dispersion Profile}
From their sample of 33 stars, M98 reported {\em central} velocity
dispersions ranging from (8.6$\pm$1.2)\,km\,s$^{-1}$ to
(9.2$\pm$1.6)\,km\,s$^{-1}$, which were obtained using several
model-independent estimators.  This is slightly lower than our global
estimate from eq.~1, but consistent within the uncertainties.

We determined the velocity dispersion profile using the maximum
likelihood method outlined in Kleyna et al. (2004).  The data were
binned such as to maintain a constant number of stars per bin, where
we chose the binsize such that no fewer than ten stars were included. A
Gaussian velocity distribution around the single mean velocity of the
entire ensemble is then assumed for each radial bin, convolved with
the observational errors.  The member velocity distribution centred on
the systemic velocity is then combined with an interloper
distribution, $P_{int}(v)$, contributing a fraction $f_{int}$ to each
bin. This allows the probability distribution of the true dispersion
$\sigma$ in the bin through an extension of eq. 1:
  
\begin{eqnarray}
p\,(\{ v_i,\sigma_i \} | (\left<v\right>,\sigma)\,)\,&\propto&
\, \Pi_i\left[ \,(1 - f_{int})\,
\frac{1}{\sqrt{2\pi(\sigma^2\,+\,\sigma_i^2)}}\,\nonumber\right.\\
&\times&\left.\,\exp\left\{-\frac{(v_i\,-\,\left<v\right>)^2}{2\,(\sigma^2\,+\,\sigma_i^2)}\right\} 
+ f_{int}P_{int}(v)\right]
\end{eqnarray}

We then perform a maximum likelihood fit over $\sigma$ and $f_{int}$,
marginalise over the interloper fraction and finally find the most
likely dispersion from the marginalised distribution.

We take account of the effects of the Galactic foreground interlopers
in our data set by assuming various associated distributions
$P_{int}(v)$ over the whole range of interest around the systemic
velocity of Leo\,I.  Consequently, we tested power-law velocity
distributions with varying indices against the uniform case and
against each other and found from a Kolmogorov-Smirnov (K-S) test that
the resulting dispersion profiles were practically identical at a
100\% confidence level. The velocity distribution of the ten
non-member stars in our sample itself is best represented by a
power-law with an exponent of $-0.7$ (with a K-S probability of 93\%),
whereas the K-S probability that these stars' velocities are uniformly
distributed is 52\%.  We emphasise, however, that the resulting
dispersion profiles do not change with respect to the adopted
parameterisation of the underlying interloper distribution.  This is
also to be expected, given the low total number of non-members
sampled, a consequence of the overall high systemic velocity of Leo\,I
relative to line-of-sight galactic halo stars.  We will, for
convenience, use the uniform case in the following.
Typically, the peak interloper fraction for the outermost bin does not
exceed 0.1 and is compatible with zero in the inner bins.

Finally, formal error bounds were determined by numerically
integrating the total probability of the data set and finding the
corresponding 68 per cent confidence intervals.  Fig.~5 shows the
resulting velocity dispersion profiles as a function of the projected
spherical radius for different membership criteria, and separately for
the subsamples.
\subsection{Radial variations}
Given the derived uncertainties in the profiles, the dispersion
profile of LeoI can be considered to be approximately flat out to the
nominal tidal radius for the majority of the data combinations shown.
Exclusion of the 3$\sigma$-outlier in the last bin leads to only a
marginal change.  Inclusion of the data from M98 with our own data
into the maximum sample decreases the dispersion in the central
3$\arcmin$ by $\sim$2\,km\,s$^{-1}$, while the associated rebinning of
the data introduces a small drop in the dispersion beyond 10$\arcmin$,
quantifying the limited sensitivity of the binned dispersions to sample
binning.

\subsection{Absence or presence of kinematical substructure}
Localized, kinematical structures in Local Group dSphs have been
detected, specifically in the form of low velocity dispersion
(cold) spatially clumpy substructure in the UMi dSph (Kleyna et
al. 2003; Wilkinson et al. 2004).  In Sextans a low velocity
dispersion (cold) central core has been found (Kleyna et al. 2004) and
independently the presence of an off-centered kinematically distinct
subpopulation has been reported (Walker et al. 2006b).  Such distinct
cold features can be interpreted as the remains of dispersing star
clusters, with their central locations arising due to the clusters being 
dragged towards the inner regions by dynamical friction (Kleyna et al. 2003,
2004; Goerdt et al. 2006).  An alternative proposal involves the after
effects of a merger between a dSph and an even smaller system (Coleman
et al. 2004).

Despite the overall flatness of Leo\,I's velocity dispersion towards
larger radii, the profile does show some radial structure.  A
noteworthy radial feature is an apparent rise in the dispersion
profile at $\sim 3\arcmin$ (r/r$_{tid}\sim0.25)$.  This trend is
persistent, regardless of radial binning and the chosen sample, and it
is also inherent in the profile observed by Mateo (2005). The major
difference between our profile and the one reported in Mateo (2005) is
that his dispersion rises continuously out to 10$\farcm$5 after
having reached its minimum after the bump.  Hence, it cannot be
excluded {\em a priori} that the local dispersion maximum may in fact
be localized and physically real.

In order to test whether our data set supports the presence of 
any underlying localized kinematical substructure,  we estimate the position-dependent 
dispersion using a non-parametric approach. 
We define the locally weighted, average dispersion as
\begin{equation}
\sigma\,(x,y)\,=\,\frac{\sum_{i=1}^N \left\{(v_i-\left<v\right>)^2-\sigma_i^2\right\}\,K(x,x_i,y,y_i,h)}
{\sum_{i=1}^N\,K(x,x_i,y,y_i,h)}
\end{equation}
(Walker et al. 2006b, after Nadaraya 1964; Watson 1964). Here, $v_i$
and $\sigma_i$ are as usual the observed radial velocities and
associated uncertainties of the $N$ targets at the location
($x_i,y_i$). The smoothing kernel $K$ is characterized by its
smoothing bandwidth $h$, which we adopted as variable such as to
include a fixed number $n$ of stars within 3$h$ at each location
($x,y$). For convenience in comparing our analyses, we follow the
prescription of Walker et al. (2006b) in adopting a bivariate Gaussian
kernel; hence \\ $K\propto
\exp\left\{-\frac{1}{2}\frac{(x-x_i)^2+(y-y_i)^2}{h^2}\right\}$.  The
number of nearest neighbors $n$ was chosen sufficiently high to yield
a statistically significant estimate of each local dispersion
measurement, but sufficiently low so that any real spatial information
would not be averaged out. Hence we varied $n$ from 5 to 30,
corresponding to at most about 25\% of the entire sample size.

The significance of any potential substructure in our data was then
determined via detailed Monte Carlo modeling. For this purpose, 10$^3$
random data sets were generated, where the ($x,y$) coordinates of our
targets were preserved, but the observed velocities and error
estimates were permuted with respect to their position. In this way,
the sample mean and dispersion were retained, whereas any spatial
information was dissolved. For each of these random data sets, the
spatial dispersion distribution (eq.~3) and its maxima and minima were
determined.  The significance $p_{hot}$ ($p_{cold}$) of hot (cold)
substructures at each spatial point was then defined by the fraction
of random data sets with a maximum (minimum) dispersion, which is
lower (higher) than the actual observed local dispersion $\sigma(x,y)$
(see Walker et al. 2006b).

We do not find evidence of any localized cold substructure with low
radial velocity dispersions in our data (see Fig.~6).  In both the
GMOS\,+\,DEIMOS data sets of Leo\,I and after inclusion of M98
targets, the significance of cold structures, $p_{cold}$, does not
exceed 50\% for any choice of the number of neighboring points. The
typical value of this significance lies at 30\%.

On the other hand, the presence of a physically real local maximum in
the radial dispersion profile could indicate the presence of a locally
{\em hot} structure at radii of $\sim$3--4$\arcmin$.  Although the
probabilities for an occurrence of hot substructure increase towards
the nominal tidal radius, where the data are only sparsely
distributed, the formal significances $p_{hot}$ are generally well
below the 80\% level. Hence, the apparent localized kinematical
structure is consistent with statistical fluctuations about a constant
dispersion value at the  1$\sigma$-level.
In particular, there is no significant evidence of any localized
structure at the radii in question -- at loci between 3$\arcmin$ and
4$\arcmin$, the probability of an occurrence of any dynamically hot
substructure is less than 15\% for any reasonable choice of $n$.

Since the submission of our paper, S06 and Bosler et al. (2006) 
have presented kinematic data sets for Leo\,I. An analysis 
of the combined data sets (also including our data and those 
of M98), which yields a sample of the order of 400 red giants 
velocities, will aid the investigation of the   
the full kinematical structure of Leo\,I and is currently underway. 

From our data, we conclude that the apparent rise in the velocity dispersion profile 
does not exclude the presence of any kinematical substructure, 
but it is probably not a localized feature. 

\subsection{(Non-) Influence of binaries}
The presence of a significant population of binaries in any
kinematical data set leads to an inflation of the observed velocity
distribution, or, in other words, the true line of sight velocity
dispersion of a stellar system is smaller than the observed
dispersion, as soon as a non-zero binary fraction is
considered. Moreover, the high-velocity tail of a binary distribution
drives a distribution function away from Gaussian, so that an
over-simple determination tends to produce larger errors in the
derived velocity dispersion, which in turn may reduce the statistical
significance of any real radial gradient in the dispersion profile
(see error bars in Fig.~6).
Fortunately, the effect of binary stars has been shown to be
negligible in dSphs in the past. Hargreaves, Gilmore \& Annan (1996)
could effectively show by using Monte Carlo simulations that the
velocity dispersion caused by pure binary orbits is small compared
with the generally large observed dispersions in dSphs, provided that
the ensemble of orbital parameters in the dSph stars is similar to the
distribution in the solar neighbourhood.  Likewise, Olszewski et
al. (1996) concluded from their simulations that none of the
kinematical estimates in the UMi and Draco dSphs significantly
changes under the influence of binary stars. This finding was
underscored by repeat observations of red giants in the Draco dSph
(Kleyna et al 2002), and also Walker et al. (2006a) conclude from 
their repeat observation of red giant velocities in the Fornax dSph 
that the impact of binaries on the measured velocity dispersion is 
negligible.  
These observed kinematics did not show any evidence that would support
an overall binary content larger than 40\%.  Moreover, the dynamically
significant fraction in the Draco sample amounted to less than 5\%.
In the case of Leo\,I, Gallart et al. (1999a) argued from their
modeled SFH based on deep HST color magnitude diagrams (CMDs) that it
is unlikely that the total binary fraction in this galaxy exceeds
60\%, again with a far lower dynamically significant number.

In order to explore the importance of binaries for the velocity
dispersion profile of Leo\,I, we added an additional term to eq.~2,
accounting for a binary star distribution $P_b(v)$, which is then
convolved with the observed velocity distribution, in accordance with
eq.~1 of Kleyna et al. (2002).  The binary probability distribution
$P_b(v)$ was adopted from Kleyna et al. (2002) and is based on the
velocity measurements of Duquennoy \& Mayor (1991) for a sample of
solar-type primary stars in the solar neighbourhood.  Kleyna et
al. (2002) then obtained a realistic distribution by taking into
account the giant branch binary evolution through circularization of
the orbits.

We note that the presence of a non-zero binary fraction does not alter
the derived interloper fraction, which is still consistent with zero
(less than 5\% in the outermost bin).  Fig.~7 shows velocity
dispersion profiles obtained under the assumption of different binary
fractions. The average errorbar on the velocity
dispersion measurements increases by 10\% for a binary fraction of 0.4
and by 25\% for an implausibly large $f_b$ of 0.8. Likewise, the
overall decrease of the dispersion profiles is, with less than
0.7\,km\,s$^{-1}$ deviation for 40\% binaries, well below the
0.5\,$\sigma$ level.  An increase of $f_b$ leads to a
progressively more pronounced fall off of the underlying dispersion
profiles. Nonetheless, the velocity dispersion profile assuming 40\%
binaries is consistent with that using no binaries at the 90\% confidence
level, as confirmed by a K-S test so that henceforth the effect of
binaries on our observed line of sight velocity dispersion profiles is
taken to be negligible.
\section{Isotropic mass estimates}
The simplest possible estimate of the mass profiles of dSphs uses
single-component dynamical models, in which it is assumed that the
mass distribution follows that of the visible component (``mass
follows light'', e.g., Richstone \& Tremaine 1986). In this context,
King models (King 1966) were often used to describe both the surface
density and the velocity dispersion profile.
However, the parameters describing the visible content of dSphs, i.e., 
core radius $r_c$ and velocity dispersion profile $\sigma$, are in all
cases studied in detail inconsistent with the mass-follows-light
assumption inherent in this style of analysis (Kormendy \& Freeman
2004). In the case here, where the approximately flat observed
dispersion profile of Leo\,I suggests that the stars orbit in a (dark
matter) halo which extends to radii larger than the nominal tidal
radius of the observed light distribution, this historical approach is
inappropriate.  Where sufficient data exist, two parameter dynamical
models for spherical stellar systems (Pryor \& Kormendy 1990; Kleyna
et al. 2002), additionally accounting for velocity anisotropies and a
dark matter component or non-parametric modelling schemes (Wang et
al. 2005, Walker et al. 2006a) are progressively being applied to
reproduce observed profiles, in particular out to large radii.

Where only limited data exist, sufficient to define a binned
dispersion profile but insufficient to support a full dynamical
analysis of the distribution function, an intermediate level of
analysis is appropriate, and is applied here. We pursued a simple
approach to obtain an estimate of the galaxy's mass and density
profile by integrating Jeans' equation (Binney \& Tremaine 1998,
eqs. 4-54 ff.) first under the assumptions of an isotropic velocity
distribution and spherical symmetry, and below considering anisotropic
distribution functions, using smooth functional fits to
represent the light distribution and dispersion profile of Leo\,I.
 
In the case of an isotropic velocity distribution, the Jeans equations 
give rise to the simple mass estimator
\begin{equation}
M(r)\,=\,-\frac{r^2}{G\,\nu}\,\left(\frac{{\rm d}\,(\nu \overline{v_r^2})}{{\rm d}\,r}\right),
\end{equation}
where $\nu$ and $\overline{v_r^2}$ denote the deprojected, three
dimensional light density distribution and radial velocity
dispersions, respectively. Both these quantities are obtained via
direct deprojection of the observed surface brightness, $I(r)$,
and projected dispersion profile $\sigma(r)$ adopting a convenient
functional form, which in this case is a Plummer model (Wilkinson et
al. 2006a, after Binney \& Tremaine 1987).
In this case, the surface brightness and associated, deprojected 3D profiles 
read 
\begin{eqnarray}
I(r)\,&=&\,\frac{I_0\,a^4}{(a^2\,+\,r^2)^2},\nonumber\\
\nu(r)\,&=&\,\frac{3\,I_0\,a^4}{4\,(a^2\,+\,r^2)^{5/2}}.
\end{eqnarray}

The surface brightness profile of Leo\,I, which we adopted from Irwin
\& Hatzidimitriou (1995), is well fit by a Plummer model (left panel
of Fig.~8) with a scale radius $a$ of $4\farcm6$, which also nicely
accounts for objects near the nominal tidal radius at $12\farcm6$.
Since the observed velocity dispersion profile, on the other hand,
appears to be consistent with being flat, we fit this component with a
Plummer profile with a large scale radius of 74$\arcmin$ to ensure a
flat shape throughout the observed radius (right panels of Fig.~8).
However, as Fig.~8 indicates, this argument is slightly sensitive to
the value of the outer bin, which may give rise to a fall-off in the
profile. Hence, we also chose a Plummer profile with a smaller radius
(36$\arcmin$), which fits the observations best. In either case, the
central value of the velocity dispersion $\sigma_0$ was determined as
9.8\,km\,s$^{-1}$.  These values hold only for the combined GMOS +
DEIMOS sample. The inclusion of M98's data leads to a decreased
central velocity dispersion and consequently yields a lower central
density $\rho_0$. The
resulting density profiles in Fig.~9 fall off faster in the case that
Leo~I has a falling dispersion profile. Either density profile reaches
a $r^{-1}$ slope at $\sim 3\arcmin$, but under the assumption of a
flat velocity dispersion the density converges to the $r^{-2}$-law
(linearly rising mass), whereas it tends to be consistent with a
$r^{-2.5\dots-3}$ behavior towards the nominal tidal radius for the
falling dispersion profile.

As discussed above our data suggest a rising feature in our velocity
dispersion profile at approximately 3$\arcmin$. Although it has been
shown above that this is not a statistically-significant localized
structure, it may be physically real, and so provide useful
constraints on the dark-matter mass distribution in Leo\,I.
Consequently, we consider this particular shape by fitting the
dispersion profile with an additional asymmetric Gaussian component of
the form
\mbox{$\sigma_{bump}\,\propto\,10^{-\alpha\,r}\,\times\,\exp\,(-10^{(r-\mu)/p})$}
overlaid on the Plummer profiles. The free parameters $\alpha$, $\mu$
and $p$ are then determined in a least-squares sense.
Using this best-fit representation in the dynamical calculations
results in an unphysical behaviour in the resulting mass profile,
i.e., a drop in the cumulative mass and the density profile at the
location of the bump (left panels of Fig.~9).  This is attributed to a
too steep rise in the additional component as opposed to the smooth
decline towards larger radii. Hence, the associated gradient, which
enters the Jeans' equation, overwhelms the gradient of the declining
portion in the dispersion and light profiles leading to a negative
contribution in the derived mass profile.  We note that varying the
parameters of the Gaussian peak, retaining consistency with the GMOS +
DEIMOS profile, does not remove this inconsistency.
Conversely, a comparable fit to the GMOS + DEIMOS + M98 data does not
exhibit this unphysical outcome, since for the respective best-fit
profile the gradient in the innermost regions does not counteract the
gradient of the more gradually declining profile at larger radii.  We
conclude that the failure of such a basic functional approach to
simultaneously yield a physically expedient representation of all
profiles considered suggests that we in fact do not see a real
substructure.  If, on the other hand, the dispersion profile could be
modeled sufficiently well in either case and the radial feature was
assumed to reflect a real structure, its particular shape would result
in a close to uniform density profile at central radii.  This would be
indicative of a central core in Leo\,I. We will explore the
plausibility of cored mass profiles in the next Section.

The upper and lower mass- and central density limits were finally
derived by determining a profile that reproduced the observed velocity
dispersion profile within the measurement uncertainties on either
side.
An estimate of Leo\,I's mass thus yields
$M_{tot}=(8\pm2)\,\times\,10^7\,M_{\odot}$ enclosed within the King 
radius of 12$\farcm$6, corresponding to 0.9 kpc at the distance
of Leo I.  The exact values depend on the choice of the formal
representation of the line-of-sight velocity dispersion profile and
are detailed in Table~4.  Similarly, the central density is found to
be $\rho_0=(2.8\pm 0.8)\,\times\,10^8\,M_{\odot}\,$\,kpc$^{-3}$, 
which is, in the light of the
measurement uncertainties, consistent with the value of $(3.4\pm
0.9)\,\times\,10^8\,M_{\odot}\,$\,kpc$^{-3}$ obtained by M98 from
single-component King fitting.
The available dynamical mass estimates for dSphs yield total masses
out to the radial limits of the respective kinematic data in the range of
3--8$\,\times\,10^7\,M_{\odot}$ (e.g, Mateo 1998, Wilkinson et
al. 2004; Kleyna et al. 2004; Chapman et al. 2005; Walker et
al. 2006a; Wilkinson et al. 2006a).  This places Leo\,I in the upper
range of known LG dSph masses.
Adopting a total luminosity of
$L=(3.4\pm1.1)\,\times\,10^6\,L_{\odot}$ (Irwin \& Hatzidimitriou) we
find an isotropic mass-to-light ratio of
$(M/L)_{tot}=(24\pm6)\,(M/L)_{\odot}$ (see Table~4). This value is
three times larger than the central value quoted by M98, who obtained
their estimate based on analyses of the galaxy's core region and
adopted a higher luminosity of $L=4.9\,\times\,10^6\,L_{\odot}$. If
we adopt this latter value for $L$, we obtain a $(M/L)_{tot}$ of
17$\,(M/L)_{\odot}$. Moreover, S06 derive an estimate of 5.3\,$(M/L)_{\odot}$, 
again with an underlying higher total luminosity of $7.6\,\times\,10^6\,L_{\odot}$, and based on 
core fitting and their value of the central velocity dispersion. 

In Fig.~10 we plot the V-band mass to light ratios versus V-band
luminosities of the known Galactic dSphs, including some of the
recently discovered systems, like Ursa Major (Willman et al. 2005) and Bo\"otes 
(Belokurov et al. 2006a), 
plus two of the M\,31 companions with published masses.  
In
particular, we adopted mass estimates from Mateo (1998) for Leo\,II
and Sculptor, whereas more recent measurements were used for the other
satellites (C\^ot\'e et al. 1999 for And\,II; Wilkinson et al. 2004
for UMi and Draco; Kleyna et al. 2006 for Ursa Major; Chapman et
al. 2005 for And\,lX; Wang et al. 2005 for Fornax; 
Wilkinson et al. 2006a for Carina and Sextans, and 
 Mu\~noz et al. 2006b for Bo\"otes).

The idea that dSphs may be dominated by dark matter out to large radii
raises the intriguing question whether all these dwarf galaxies could
be enclosed in comparable dark matter halos of similar total mass, as 
first pointed out by Mateo et al. (1993). This feature would then argue in favor of
an intimate relation between the dark matter halos of the dSph systems. 
Moreover, cosmological simulations suggest that it is
possible that there is a minimum mass for a dark matter halo that
contains stars (e.g., Read et al. 2006a and references therein). It
is also an interesting question to ask whether all the dSphs have
similar mass haloes, as this might tell us something about the
properties of the dark matter. 
Under such an assumption, the halos can be represented by the the simple relation
\begin{equation}
(M/L)_{tot}=(M/L)_{\ast} + M_{DM}/L,  
\end{equation}
where $(M/L)_{\ast}$ is the intrinsic mass-to-light ratio of the
stellar component in the galaxy, $M_{DM}$ denotes the mass of the dark
matter halo and $L$ is the integrated luminosity in the V-band (M98;
Wilkinson et al. 2006a).  It is realistic to assume that the
(M/L)-ratio of the the luminous component in dSphs is of the order of
the mean value observed in low-concentration globular clusters, which
are characterised by low values implying no dark matter (Pryor et
al. 1989; Moore 1996; Dubath \& Grillmair 1997; M98; Baumgardt et
al. 2005).  We do not correct the observed M/L-ratios for the (small)
effects of stellar evolution, accounting for each individual dSph's
SFHs. Thus we set $(M/L)_{\ast}=1.5\,(M/L)_{\odot}$.  As the best fit
dashed line in Fig.~10 implies, the observed dSphs scatter around the 
expected relation for a population of luminous objects that are
embedded in a dark halo with a common mass scale of the order of
3$\,\times\,10^7\,M_{\odot}$.  Considering the generally large scatter
in such a (M/L)-$M_V$-diagram, Leo\,I fits well into the global
picture and apparently is governed by the same uniform dark halo
properties. 
 It is hence worth noticing that Fig.~10 demonstrates that the narrow range of dSph
velocity dispersions of 6--10\,km\,s$^{-1}$, combined with their
rather similar physical length scales, can generally be interpreted in
terms of a common mass scale so that this kinematical piece of
information can reliably be used as a proxy for an overall
underlying mass distribution (see also Wilkinson et al. 2006a).

\section{Velocity anisotropy}
The mass estimates derived above are based on the assumption of an
isotropic velocity distribution, i.e., the velocity anisotropy
parameter $\beta=1-{\left<v^2_{\theta}\right>}/{\left<v^2_r\right>}$
was assumed to be zero. However, most of the kinematic studies of
dSphs have revealed that a non-negligible amount of anisotropy is
required to account for the shapes of the galaxies, and the observed
dispersion profiles (\L okas 2001, 2002; Kleyna et al. 2002; Wilkinson
et al. 2004).  Since the neglect of a non-zero anisotropy has the
effect of under- or overestimating the dark halo mass, we also shall
consider this problem by solving the Jeans equation
\begin{equation}
\frac{{\rm d}\,(\nu\overline{v^2_r})}{{\rm d}\,r}\,+\,2\beta\,\frac{\nu \overline{v_r^2}}{r}\,=\,
-\nu\,\frac{{\rm d}\,\Phi}{{\rm d}\,r} 
\end{equation}
for a varying $\beta$.  Here, $\nu$ and $\overline{v_r^2}$ denote
again the 3-dimensional light and dispersion profiles, and $\Phi$ is the
gravitational potential associated with the underlying mass
distribution $M(r)$. The actual observable quantity, the line-of-sight
velocity dispersion $\sigma$, is then obtained by direct integration
along the line of sight, and under the constraints
$\rho\,\overline{v_r^2}\rightarrow0$ for $r\rightarrow\infty$ and
$\beta=$\,const. This gives rises to the one-dimensional expression
\begin{eqnarray}
\sigma^2(R)\,  =  \,\frac{2G}{I(R)}\,\int_R^{\infty}\,&{\rm d}x&\,\nu(x)\,M(x)\,x^{2\beta-2}\nonumber\\
\times\,\int_R^x\,&{\rm d}y&\,\left(1-\beta\,\frac{R^2}{y^2}\right)\,\frac{y^{-2\beta+1}}{\sqrt{y^2-R^2}} 
\end{eqnarray}
(Binney \& Mamon 1982; \L okas \& Mamon 2003; Wilkinson et al. 2004).
By adopting analytical prescriptions for the involved functions, one
can then proceed to derive $\sigma$ numerically.

In concordance with the previous section, we chose to describe Leo\,I's surface brightness 
profile $I(R)$ and the associated 3D-deprojection with a Plummer-profile 
using the 
best-fit parameters to the observations. 
We then use two different representations of the mass profile $M(r)$, which enters this formalism, 
spanning the range of plausible mass distributions. 

\subsection{NFW halo mass distribution}

Based on high-resolution cosmological simulations, Navarro, Frenk \&
White (1995, hereinafter NFW) demonstrated that the density profiles
of dark matter halos are well fit by a simple function with a single
free parameter, the characteristic density. This solution has been
applied to a wide range of halo masses, ranging from galaxy clusters to
small scales such as the cosmological dark matter halos associated
with the dSphs.  Upon transformation of the variables, the density
profile corresponds to a mass profile of the form
\begin{equation}
\frac{M(s)}{M_v}\,=\,g(c)\,\left\{\ln (1+cs)\,-\,\frac{cs}{1+cs}\right\}, 
\end{equation}
where $s=r/r_v$ denotes the radial distance in units of the virial
radius and $M_v$ is the mass enclosed within the virial radius. The
latter is generally identified with the total mass of the
halo. Finally, the concentration parameter $c$ is used to describe the
shape of the profile and defines the amplitude function
\begin{equation}
g(c)\,=\,\left[\ln(1+c)\,-\,c/(1+c)\right]^{-1}.
\end{equation}
For a detailed discussion of these parametrizations and the model, we
refer the reader to NFW or \L okas \& Mamon (2001).  Since the
NFW-profile has the unphysical disadvantage of diverging at large
radii, it is convenient to consider only radii within a certain
cut-off radius. We follow long-standing practice and assume this
cut-off to coincide with $r_v$; beyond this point, the overall density
distribution becomes unreliable in most natural cases (e.g., \L okas
\& Mamon 2001).

The concentration $c$ has been shown to scale with mass. Here we adopt
an extrapolation of the formulae derived in the $N$-body simulations
for $\Lambda$CDM cosmology of Jing \& Suto (2000) to the small masses
of dwarf galaxies.  Hence,
\begin{equation}
c\,=\,10.23\left(\frac{h\,M_v}{10^{12}\,M_{\odot}}\right)^{-0.088}, 
\end{equation}
where $h$ denotes the Hubble constant in units of
100\,km\,s$^{-1}$\,Mpc$^{-1}$.  In concordance with current
cosmological results, we will use $h$=0.7 for the remainder of this
work.
Likewise, the virial radius is related to the virial mass via 
\begin{equation}
r_v\,=\,\quad206\left(\frac{M_v}{10^{12}\,M_{\odot}}\right)^{1/3}\qquad \left[kpc\right] 
\end{equation}
This leaves $M_v$ and the anisotropy $\beta$ (assumed to be constant) 
as the only free parameters to be 
determined in a fit of the NFW model profile to the observations.

As Fig.~11 (top panel) implies, none of 
the values for the anisotropy which we have considered yields an 
adequate fit of the overall dispersion profile. It rather appears that 
a progressively radial anisotropy is needed to explain the 
observations at larger radii, whereas the inner parts of the velocity 
dispersion are best represented by an isotropic velocity tensor.  It 
is worth noticing that the innermost bin, with a smaller value for the 
dispersion in the sample after inclusion of the M98 data points, may 
suggest the presence of some tangential anisotropy in the innermost 
regions. However, based on the observations,  $\beta$ is unlikely 
to fall below $-0.5$.  For the four different $\beta$ curves that we 
tested (i.e., $-$0.5, 0, 0.5 and 1), we obtained reduced $\chi^2$ 
values of (0.2, 0.16, 0.22, 0.78).  The final best fit in terms of a 
minimized $\chi^2$ was achieved for $\beta$\,=\,0.05 so that the 
overall profiles are in agreement with an isotropic velocity 
distribution.  The resultant best-fit total virial mass is 
$M_v$\,=\,1.0 $\times 10^9\,M_{\odot}$,  with a corresponding virial radius 
of $r_v =21$\,kpc and a concentration $c=19.4$. 
 Furthermore, the mass within 
the nominal observed tidal radius amounts to 7.7$\times 
10^7\,M_{\odot}$. It is, unsurprisingly given the near-isotropic 
distribution deduced, in good agreement with the value obtained under 
the assumption of a purely isotropic velocity distribution.

\subsection{Cored halo profile}
The density profile that we obtained under the assumption of velocity
isotropy appears to be indicative of a flat, close to uniform density
core at small radii (see bottom panels of Fig.~9). In conjunction with
the observational indications of cored density profiles in low-luminosity
galaxies and in particular in some of the dSphs (\L okas 2002; Kleyna
et al. 2003; Strigari et al. 2006; Wilkinson et al. 2006a) this
prompted us to also consider this particular density distribution in
our calculations of Leo\,I's dispersion profile (eq.~7).
We parameterize the cored density profile in terms of a generalized
Hernquist profile (Hernquist 1990; Zhao 1996; Read \& Gilmore 2005) as
\begin{equation}
\rho(r)\,=\,C\,\left(\frac{r}{r_s}\right)^{-\gamma}\,\left[1+\left(\frac{r}{r_s}\right)^{\alpha}\,
\right]^{(\gamma-3)/\alpha}
\end{equation}
where $C$ is a normalization constant, $r_s$ is a scale radius,
$\gamma$ is the log-slope of the density profile within $r_s$ and
$\alpha$ determines the smoothness of the transition towards the
density profile at larger radii.
The case of $\alpha\equiv1$ has often been discussed and fit to
dispersion profiles in the literature, since it allows one to envisage
a broad class of dark matter halo representations (Jing \& Suto 2000;
\L okas 2002).  A slope of $\gamma>0$ then corresponds to the cuspy
halos, where NFW-like profiles as discussed above are represented by
$\gamma=1$. The opposing class of models with a cored halo is realized
by $\gamma=0$.

It turns out that a least squares fit under the constraint
$\alpha=1$, yields an unsatisfactory representation of our
observations. Thus we chose to use the profile according to eq.~13 in
our fits, leaving $C$, $r_s$, $\alpha$ and $\gamma$ as free
parameters.  The resulting parameters of $r_s=0.3$\,kpc, $\alpha=1.6$
and $\gamma=0$ improved the reduced $\chi^2$ from 55 (3-parameter fit)
to 13 (4 parameters).  It is worth noticing that indeed the best fit
was obtained for a pure core profile ($\gamma\equiv0$).
Fig.~11 (bottom panel) displays the resulting radial velocity
dispersion profiles for varying degrees of (constant) anisotropy.

As for the case of the cuspy dark halo treated above, we find that our
observed velocity dispersion is broadly consistent with an isotropic
velocity distribution,where the best fit yielded a $\beta$ of 0.05.
Still, the observations allow for a wider range in the anisotropy,
from $-$0.5 to 0.5.  In this context, a slight amount of tangential
anisotropy can account for the shape of the profile toward
the inner regions, whereas significant radial anisotropy is ruled out at all
radii. The values for the associated reduced $\chi^2$ statistics for
the cored halo are (0.39, 0.25, 0.40, 1.94) for respective $\beta$ values 
of ($-$0.5, 0, 0.5, 1).

Finally, we also adopted a radially varying anisotropy in our calculations, 
where we parameterized $\beta$ according to the traditional prescriptions 
by Osipkov (1979) and Merrit (1985) as 
$ \beta_{\rm OM}(r)\,=\,{r^2} / ({r^2+r_a^2}) $, 
where $r_a$ denotes the anisotropy radius, at which the transition from isotropic to 
radial orbits occurs (see also \L okas 2002). 
Under the assumption of a cored density profile and this particular $\beta(r)$, we 
can fit our observed line-of-sight velocity dispersion profile fairly well 
(see the thick solid line in Fig.~11, bottom panel). The best-fit (at a $\chi^2$ of 
0.02) is then obtained for an anisotropy radius of $r_a=0.13$\,kpc. 
In particular, the resulting curve is able to account for the observed rising shape 
of the profile discussed in Sect.~5.2, albeit by requiring 
a considerable amount of radial anisotropy at relatively small radii. 

Another possible explanation of the dispersion profile of Leo\,I 
might be the presence of two stellar components with different length-scales 
and associated different velocity dispersions. McConnachie et al. (2006)
have recently presented dispersion profiles derived for such a system. 
Their profiles bear a striking resemblance to that of Leo\,I and 
might account for the initial rise in the projected dispersion. 

In concluding we note that the enclosed mass within the tidal radius 
amounts to 8.5$\times 10^7\,M_{\odot}$ under the assumption of a cored halo in Leo\,I. 
This value is larger than under the assumption of a NFW-like mass distribution, but 
still in agreement with the numbers derived from the Jeans equations above.
\section{The Metallicity of Leo\,I}
The observed wavelength region in the near-infrared and our spectral
resolution, coupled with the S/N of our observations also enabled us
to get an estimate of the red giants' metallicites from the equivalent
widths (EWs) of the Ca triplet (CaT).
This procedure succeeded for 58 of the radial velocity member stars
targeted with GMOS, all of which lie well within the nominal tidal
radius.  For the remainder of our targets, the S/N did not allow us to
reliably measure EWs, where a lower limit for these measurements was
S/N$\sim 10$.

\subsection{Calibration of the metallicity scale}
The prominent \ion{Ca}{2} feature may be calibrated onto global
metallicities since the linestrength is a linear function of
metallicity for red giants. By defining a reduced width $W'$, which
accounts for the luminosity difference between the targeted star and
the horizontal branch (HB) of the system, one can effectively remove
any dependence on stellar gravity, effective temperature, and
distance.  Thus, the CaT has become a well defined calibrator for
assessing the metallicity for old, globular cluster-like populations
(Armandroff \& DaCosta 1990, Rutledge et al. 1997a,b).  Consequently,
the derived line widths are generally calibrated onto reference scales
for globular clusters of known metallicities (Zinn \& West 1984;
Carretta \& Gratton 1997; Kraft \& Ivans 2003).  Leo\,I has a
prominent old population (e.g., Held et al.\ 2000), but its dominant
populations are of intermediate age (e.g., Gallart et al.\ 1999a,b).
Nonetheless, we can still apply the CaT technique, since Cole et al.\
(2004) have extended its calibration to younger ages.  These authors
demonstrated that the CaT technique can be used within an age range
from 2.5 to 13 Gyr and for a metallicity range spanning $-2 <$ [Fe/H]
$< -0.2$.
Since our Leo\,I observations aimed primarily at measuring accurate
radial velocities, we did not target any globular clusters, which
could have been used as calibration standards for the CaT
technique. Hence, we had to rely on standard calibrations devised in
the literature.  Following long standing practice, we employed the
definition of Rutledge et al. (1997a, hereafter R97a) for the
linestrength of the CaT as the weighted sum of the EWs, giving lower
weight to the weaker lines:
\begin{equation}
\Sigma W\,= \,0.5\,W_{8498}\,+\,W_{8542}\,+\,0.6\,W_{8662}.
\end{equation}
For a few of the spectra the Ca line at 8498\AA~was too weak to be
accurately measured or was hampered by too low a S/N over the respective
bandpass. In this case we used a relation derived from the high S/N spectral measurements of a
sample of Galactic globular clusters from Koch et al. (2006a):
\begin{equation}
\Sigma W\,=\,1.13\,(W_{8542}\,+\,0.6\,W_{8662})\,+\,0.04.
\end{equation}
This relation was applied to the low-quality spectra where only the
two strongest lines had S/N$>$10.
R97a defined the reduced width as 
\begin{eqnarray*}
W'  &=&  \Sigma W  + 0.64\,(\pm 0.02)\,(V-V_{HB})\,,
\end{eqnarray*}
where $\Sigma W$ denotes the CaT linestrength and V$_{HB}$ is the
magnitude of the horizontal branch of Leo I.  For Leo\,I, the HB
luminosity is (22.60\,$\pm$\,0.12)\,mag according Held et al.\ (2001),
who measured it based on this galaxy's RR\,Lyrae population.

The final calibration of the reduced width in terms of stellar
metallicity is then obtained via the linear relation
\begin{eqnarray*}
\left[Fe/H\right]_{CG} &=&  -2.66\,(\pm 0.08) +0.42\, (\pm 0.02) W'\,
\end{eqnarray*}
(Rutledge et al. 1997b), where we chose to tie our observations to the
reference scale of Carretta \& Gratton 1997, hereafter CG97) unless
stated otherwise.

The EWs of the calcium lines were determined in analogy to the methods
described in Koch et al. (2006a), i.e., by fitting a Gaussian plus a
Lorentzian profile across the bandpasses defined in Armandroff \& Zinn
(1990) and summing up the flux in the theoretical profile.  The median
formal measurement uncertainty achieved in this way, incorporating EW
measurement and calibration errors, is 0.12\,dex.  However, we note
that the intrinsic accuracy of the CaT based metallicity measurements
may be significantly lower. For instance, variations
in the HB level of a composite stellar system with both age and
metallicity can give rise to a systematic error of the order of 
0.05--0.1\,dex (Cole et al. 2004). Moreover, 
the general incompatibilitiy between the
[Ca/Fe] ratio in dSphs and the Galactic globular clusters adopted as 
calibrators is another source of uncertainty. High resolution 
abundance studies of red giants in dSphs have shown that their 
$\alpha$-element abundance ratios lie significantly below those 
in Galactic halo stars at comparable metallicities (e.g., Venn et al. 2004), 
and also 
the two red giants observed in Leo\,I by Shetrone et al. (2003) 
follow this trend of $\alpha$-deficiency. 
Moreover, the low-resolution studies of Bosler et al. (2006) underscore 
these low abundance ratios as compared to the halo. 
The systematic uncertainties 
resulting from such a priori unknown variations in [Ca/Fe] for our targets 
can reach up to 0.2\,dex (e.g., Koch et al. 2006a). It is worth noticing 
that such effects can be overcome in future studies by calibrating 
the reduced width directly onto [Ca/H] (Bosler et al. 2006), thus circumventing 
any dependence of the calibrations of the galaxy's SFH, in contrast to the 
presently employed method.  

\subsection{Metallicity distribution}
Fig.~12 shows the resulting histogram of our metallicity
estimates. The MDF is clearly peaked at a median metallicity of $-1.31\pm0.02$\,dex
on the scale of CG97, with a measurement error of 0.12\,dex 
($-1.61$ on the scale of Zinn \& West 1984;
hereafter ZW84).  This value is in excellent agreement with the study
of Bosler et al. (2006), who derived a CaT metallicitity of
$-1.34\pm0.02$\,dex (mean measurement error of 0.10) for a sample of 101 red giants 
(note, however 
that the lack of any common targets between the studies of Bosler et al. [2006] 
and ours does  not allow us to compare measurements of individual stars). 
Our results also 
agree within the uncertainties with the low-resolution studies of Suntzeff et al. (1986), 
who obtained $-1.9\pm0.2$\,dex (ZW84).  Moreover they are
consistent with the estimates from photometric analyses, which broadly
agree on a mean metallicity around $-1.6\pm0.4$\,dex (ZW84 scale; Demers et
al. 1994; Bellazzini et al. 2004)

The formal 1\,$\sigma$-width of the MDF is 0.25\,dex. Taking into
account a broadening due to the observational uncertainties, the
intrinsic metallicity dispersion amounts to 0.22\,dex. With a full
range in [Fe/H]$_{\mathrm{CG97}}$ of approximately one dex between
$-$1.8 and $-0.8$\,dex (note the most metal poor star, which we
discuss below) we observe a smaller spread than is seen in a number of
other dSphs (Shetrone, C\^ot\'e, \& Sargent 2001; Pont et al.\ 2004;
Tolstoy et al.\ 2004; Koch et al.\ 2006a), where the MDF may span more
than 2\,dex and exhibits a well populated metal poor tail.  Leo\,I has
been shown to exhibit a rather narrow RGB.  By means of detailed
modeling of Leo\,I's CMD and in particular of this narrow feature,
Gallart et al.\ (1999a) suggested that this might be indicative of the
absence of any significant chemical enrichment in this dSph.  Clearly,
the spectroscopic data show that this is not correct and that Leo\,I
underwent considerable enrichment of at least 1 dex in total range.

There is one object in the sample, \#774, for which we derived a CaT
based metallicity of $-2.61$\,dex.  There is no evidence in the
spectrum that the linestrength of this target has been underestimated,
since its S/N of $\sim$45 allowed an accurate determination of the
triplet's EWs (see Fig.~13)\footnote{We note that this metal poor
regime was not sampled by the GCs in R97b so that the determination of
such metallicities relies on an extrapolation of the existing
calibrations.}. Although weak neutral metal lines around the CaT
region do not show up clearly at our nominal spectral resolution and
are thus mostly blended, there are still some additional absorption
features discernible from the noise (see Fig.~13).  Among these are
for instance the distinct \ion{Fe}{1} blend at $\sim$8514\AA\ and a
visible \ion{Fe}{1}-line at $\sim$8689\AA. Though clearly present in
our stars with highest metallicities, these features are hardly
detectable in the continuum of the metal poor candidate. By assuming
appropriate line bandpasses around these two iron features, we employed
the same fitting technique as for the CaT described in Section~8.1 in
order to derive a rough qualitative estimate of these lines' EWs in
each of the stars, where possible.  From this point of view, \#774 is
well consistent with being a rather metal poor object, in which these
two EWs are practically governed by the continuum noise.  Furthermore,
its location well within the galactic boundary (2.5 core radii, 0.67
tidal radii, resp.), its consistency with the systemic velocity
(99.9\% confidence level) and the low interloper fraction around
Leo\,I's radial velocity make it appear unlikely that this star is a
contaminating foreground dwarf.

\subsection{Radial gradients and implications for evolutionary models}
In those dSphs of the Local Group, in which both old and
intermediate-age stellar populations have been detected, the younger
components often are more strongly centrally concentrated.  In a
number of dSphs with predominantly old populations, population
gradients have been found in the sense that red HB stars are more
centrally concentrated than blue HB stars (e.g., Da Costa et al.\
1996; Hurley-Keller et al.\ 1999; Harbeck et al.\ 2001; Tolstoy et al. 2004), 
a trend that
is mirrored also by the presumably more metal-rich red giants. However,
as also pointed out by Harbeck et al.\ (2001), this trend is not seen
in {\em all} dSphs.

Since Leo\,I hosts dominant intermediate-age populations as well as
old populations,  one may expect that the more metal-rich and
presumably younger stars should  then  also be more concentrated with
respect to a spatially extended metal-poor population.  Our data
suggest, however,  that the metal-rich and metal-poor components in
our MDF are drawn from the same spatial distribution.  This is
reflected by the comparison of the cumulative radial distributions  of
the more metal-poor ([FeH] $<1.3$\,dex) versus the metal-rich ([FeH]
$\ge1.3$\,dex) populations in Fig.~14. Via a K-S test we could
underscore this lack of a radial metallicity gradient at the 93\%
confidence level. This finding is in concordance with the lack of a
considerable population gradient in Leo\,I's CMD. As was reported by
Held et al.\ (2000), both the old HB population and its numerous
intermediate-age counterpart exhibit essentially the same spatial
distribution (see also the right bottom panel of Fig.~14).  In this
respect, Leo\,I differs from other dSphs with prominent
intermediate-age populations such as Carina (e.g., Harbeck et al.\
2001; Koch et al.\ 2006a), Fornax (Grebel 1997; Stetson et al.\ 1998) 
and Leo\,II (Bellazzini et al.\ 2005).
Moreover, we find that the kinematics of the metal poor 
and the more metal rich samples do not differ significantly, neither. 
This is also in contrast to what  is found for  
different stellar components in many other dSphs (Tolstoy et al. 2004; 
Ibata et al. 2006): While the sample with [Fe/H]$<-1.3$ exhibits 
a dispersion of 9.1$\pm$2.8\,km\,s$^{-1}$, for those stars with 
[Fe/H]$\ge-1.3$, $\sigma=10.7\pm$3.1\,km\,s$^{-1}$, which is an 
insignificant difference in the light of the uncertainties. 

This behavior may indicate that Leo\,I's star formation occurred from
a well mixed reservoir of gas that was little affected by local
accretion of material or spatially localized outflows. In this vein,
it was suggested by Held et al.\ (2000) that the size of this galaxy
has not significantly changed and that it is unlikely to have
undergone severe structural changes in the course of its evolution,
i.e., since the first onset of star formation. In such a model one
must assume that external effects such as tides, accretion or minor
mergers may not have played a significant role in Leo\,I's history, neither
kinematically nor chemically. 
On the other hand, S06 claim that 
the preferential periods of SF in Leo\,I may be linked to perigalactic passages, 
which argues in favor of tidal interactions. 
Still, without extant detailed knowledge of this dSph's orbit (via proper motions), 
this has to await further investigation.

\subsection{Comparison with simple chemical evolution models}
Present-day chemical evolutionary models agree on the fact that dSphs
are mainly controlled by low SF efficiencies compared to the solar
neighborhood, and by strong galactic winds of the order ten times the SF
rate (Lanfranchi \& Matteucci 2004).  For want of a detailed model
prediction of Leo\,I's MDF from its SFH we follow long standing
practice in applying basic models of chemical evolution to describe
the observations (e.g., Pagel 1997, ch. 8), without employing any
deeper nucleosynthetic or physical prescriptions. Such more detailed
modeling is left for future work.

In that spirit, Fig.~12 shows the predicted MDFs for two simple models
of chemical evolution.  To start with, we obtained a best-fit
representation of a modified simple closed-box model allowing for
outflows, which is exclusively parameterized by the effective yield
$p_{\mathrm{eff}}$.
Basically, this value is proportional to the true nucleosynthetic
yield and takes into account effects of the loss of metals.
Consequently, the mean metallicity is reduced to below the true yield
by the same factor, which relates the outflows and the actual SF rate.
A low yield of 0.06\,$Z_{\odot}$ is well able to reproduce the metal
poor mean location of the peak of the MDF, when compared to the solar
neighborhood, which peaks at around $-0.2$\,dex (Nordstr\"om et
al. 2004 and references therein).  However, this prediction tends to
overestimate the number of stars below $-1.8$\,dex (note that the most
metal poor object in our sample was excluded from the fit), leading to
the well-known G-dwarf problem.  
We note that a comparison of these models, which are based 
on predictions for long-lived stars, to our observations of K-giants with predominantly 
negligible lifetimes is still valid,  since the simple model of chemical evolution 
does qualitatively predict an excess of metal poor K-giants under the assumptions 
of a standard IMF and loss of metals from the galaxy (Koch et al. 2006b).  
Moreover, it has been shown that there is no noticeable difference between 
the K-giant MDF in the solar neighbourhood (McWilliam 1990) and 
the local G-dwarf MDF from, e.g., Nordstr\"om et al. (2004). Hence we 
carry on with the long-standing practice of applying the G-dwarf 
predictions to K-giant MDFs as tracers of low-mass stars.

After allowing for an admixture of a
non-zero initial metallicity (prompt initial enrichment, PIE), the
overprediction toward the metal poor tail is reduced and it turns out
that a pre-enrichment of $Z_0\,=\,0.013\,Z_{\odot}$ yields a
remarkably good representation of the overall shape of the observed
MDF.  We refer the interested reader to Koch et al. (2006a, 2006b) for a
discussion of these standard models.

\section{Summary and Conclusions}
We have obtained radial velocity measurements for 120 red giants in
the Leo\,I dSph.  These data not only increase existing published samples by
more than a factor of three, but also extend in radial distance beyond
previous studies of the central regions of the galaxy (M98).

Although the weak velocity gradient that our data exhibit coincides
with an apparent rotation about the observed major axis of Leo\,I, the
associated amplitude has been shown to be insignificant.  
Moreover, we do not detect any trend of mean radial velocity 
with major axis distance, and in particular we do not find any evidence of an 
asymmetric velocity distribution towards the outer regions, in the sense that 
the eastern (western) halves of the galaxy would preferentially exhibit 
an excess of low (high) velocity outliers outside the formal tidal boundary.  
Supplemented
by Leo\,I's high radial velocity, which we confirmed, and its large
distance from the Milky Way, Leo\,I most likely represents an isolated
system, which is currently not affected by Galactic tides, at least from 
a kinematical point of view.  This 
is also consistent with our detection of nine radial velocity members
outside the nominal `tidal' radius, and implies that the difficulty of
the King-model functional fit to surface brightness data at some
(outer) radius is not necessarily a dynamical evidence of a physical effect. `Tidal'
radii for dSphs, as typically derived from photometry, are labels, and are not
proven to be related in any way to gravitational tides.  The results
of S06 indicate that Leo\,I's surface brightness
profile exhibits multiple breaks at large radii, where it deviates
from common King models, while  
further recent photometric analyses do not provide any such
clear evidence of photometric complexity above the noise level 
(Smol\v ci\'c et al. 2006).
Still the 
detection of dSphs' stellar populations far out beyond 
the limiting radius, as in the case of Carina (Mu\~noz et al. 2006a),  
suggests that some dSphs may in fact be perturbed systems, 
at least in their outermost regions. Clearly this has implications 
for estimates of the dark matter content of dSphs at the largest 
radii, as tidal effects may inflate the projected dispersion in these 
regions. 

Our derived projected radial velocity dispersion profile is flat out
to the nominal tidal radius, a result which is turning out to be a 
very common feature among the dSphs of the MW system, being established
in numerous studies of large radial velocity samples out to and even
beyond their nominal tidal radii (Mateo 1997; Kleyna et al.  2001,
2003, 2004; \L okas 2001, 2002; Wilkinson et al. 2004; 2006a,b;
Mu\~noz et al. 2005; Walker et al. 2006a; Westfall et al. 2006).
Coupled with the overall high (M/L) ratios, and lack of unambiguous evidence for
non-equilibrium effects, the prime explanation of flat dispersion
profiles is that dSphs are dark matter supported at all radii,  
although  the N-body simulations of S06 indicate that 
also tides could  reproduce such features in Leo\,I. 

Nevertheless, at its large present Galactocentric distance, tidal influences would
be plausible only if Leo\,I's orbit were close to radial. However, our
current data set does not allow us to constrain its proper motions and
any information about its orbit will have to await future accurate
proper motion measurements. 

It turns out that the observed dispersion profiles are largely
consistent with an isotropic velocity distribution, but non-negligible
amounts of radial (tangential) anisotropy may be needed to account for
the shape of the dispersion profile at the largest (smallest)
distances.

An apparent rise in the velocity dispersion profile at nearly the core
radius was demonstrated not to be a localized dynamical substructure
in the galaxy.  By using simple prescriptions to describe the shape of
the observed velocity dispersion profile we obtain an estimate of the
enclosed mass within the nominal `tidal' radius of 900\,pc of
$\sim$8$\times 10^7M_{\odot}$ under the assumption of velocity
isotropy.  Such a mass and the resultant high mass-to-light ratio of
$\sim$24\,($M/L$)$_{\odot}$ are higher than previously derived
values (M98; S06). From their ($M/L$) of $\sim$6--13\,($M/L$)$_{\odot}$, M98
concluded that Leo\,I contains a significant dark component.  These
estimates are in reassuringly good agreement with the values derived
from light profiles obtained from high-quality photometric data under
the assumption of an extended constant dark matter halo density
(Smol\v ci\'c et al. 2006). 
Hence, our results strengthen this view and are in concordance with an
extended dark matter distribution with a halo mass of a typical scale
($\sim$3$\times 10^7M_{\odot}$), which is shared by the other dSphs of
the LG. 

By adopting a NFW density profile for Leo\,I, we estimate its total
{\em virial} mass within its dark matter halo to be $\sim$1 $\times
10^9 M_{\odot}$. The mass within the tidal radius is estimated to be
$\sim$7.7 $\times 10^7 M_{\odot}$.
This does not change much under the assumption of a constant density
core in Leo\,I, where the enclosed mass and best-fit anisotropy are
consistent with the values from a cusped profile.  Although the density profile 
obtained using Jeans equations, under the assumption of velocity isotropy,  
indicates a flat central
core, this scenario is not significantly preferred against the NFW
case if one exclusively argues on the base of the velocity dispersion
profile.
In general, the shape of the observations seems to indicate that dark
matter in fact dominates this dSph at large and small distances.
We are aware that analysis only of a dispersion profile $\sigma (r)$
is not sufficient to derive a unique kinematical distribution function
of a stellar system. Unless more numerous data sets and 
in particular the higher moments of the velocity
distribution are considered, there remains a degeneracy between
anisotropy and density, leading to identical line-of-sight dispersion
profiles (e.g., Merrifield \& Kent 1990; \L okas \& Mamon 2003).
While in general such an analysis requires more data than is currently
available for Leo\,I, this will be pursued in future papers dealing
with more detailed dynamical modeling (Wilkinson et al., in
preparation).

We derived metallicites for 58 of the member stars from the well
established CaT calibration.  In compliance with previous estimates
that were drawn from both photometric and low-number spectroscopic
analyses we find a mildly metal poor mean of $-1.31$\,dex (CG97 scale)
and a full spread in metallicity covering 1\,dex. This is noticeably
smaller than the ranges found in dSphs with extended or complex SFHs
(Grebel 1997; Pont et al. 2004; Tolstoy et al. 2004; Koch et
al. 2006a).  This spread exceeds the predictions from CMD modeling if
one accounts for the narrow RGB and Leo\,I's existing intermediate-age 
population.  
Hence, there may be also a large range
in ages present in Leo\,I, as hinted at by the presence of also an
old population (Held et al. 2000, 2001) and the preliminary age
determinations by Bosler et al. (2004).
Practically all dSphs that contain significant intermediate-age
populations do show a central concentration of their young components
(Stetson et al. 1998; Harbeck et al. 2001; Bellazzini et
al. 2005). Intriguingly, Leo\,I's prominently younger and presumably
metal rich population does not exhibit this distinct population
gradient. Neither its stellar populations (Held et
al. 2000) nor the metallicities we derived here show any evidence of radial gradients. 
 In this respect,
Leo\,I is rather unusual among these systems.
This property may again point to a negligible role of external
influences, such as localized accretion of material, in Leo\,I's
evolution, allowing for good mixing of its gas. Since tides
would affect this mixing, it is therefore possible that its current large
distance from the MW may be representative of its whole evolution and
that it never approached the Galaxy very closely. 
This view is consistent with the finding that the
fraction of intermediate-age populations is closely related to the
dSphs' Galactocentric distance (e.g., Grebel 2003 and references therein).  
This behavior can be taken as
indicative of an increased influence of (e.g., ram pressure) stripping
close to the MW (van den Bergh 1994; Grebel 1997; Grebel et al. 2003)
and will not have affected Leo\,I to a major extent.
We note, however, that S06 argue that the presence of the various 
stellar populations in Leo\,I may in fact be linked to much closer perigalactic passages, 
which facilitated the onset of SF, a scenario that would then 
incorporate a non-negligible role of tides in Leo's evolution.

We identified one comparatively metal poor star in our sample, which is
unlikely to be a foreground contaminant. With its likely [Fe/H] of
$\sim$$-$2.6 (CG97 scale) it lies 0.8\,dex below the metal poor tail
of our observed MDF. 
It has been suggested that the explosion of only one massive pregalactic Population\,III 
star  produces a sufficient amount of energy to expel the entire gas 
from the first (high-$z$) cosmological minihalos (Bromm et al. 2003).  
If the expelled gas falls back afterwards, the subsequently forming generation of 
stars, which we would observe as the most metal poor and the oldest stars in dSphs, should bear the chemical signature of 
these Population\,III stars (e.g., Beasley et al. 2003; Kawata et al. 2006).  
This makes metal-poor stars in dSph galaxies the prime targets for searches of the signatures of 
the first stars. 
However, no star with a metallicity below $\sim-$3\,dex 
has been detected in a dSph so far (Fulbright, Rich \& Castro 2004; Sadakane et al. 2004). 
The question of whether the lack of any such extremely metal poor objects in 
the current data is  merely due to 
the low-number statistics of present-day observations, hence reflecting 
an incomplete sampling, or whether such objects generally 
do not exist in dSphs remains unsolved.  
Thus the presence of such a comparatively low metallicity object 
in our data 
provides an invaluable testing ground to study the early phases of
general galactic evolution.
\acknowledgments
We thank Mike Irwin for providing the CASU photometry, 
 Lee Clewley for support with the 
GMOS pipeline, Rachel Johnson for assistance in preparing the observations 
and Vernesa Smol\v ci\'c for providing us her photometric results before publication. 
AK gratefully acknowledges support from the European Commission under 
the Marie Curie Early Stage Research Training programme.
AK and EKG are supported by the Swiss National Science 
Foundation through grant  
200020-105260/1. 
MIW and ADM acknowledge the Particle Physics and Astronomy Research Council for 
financial support. JTK gratefully acknowledges the support of the Beatrice Watson 
Parrent Fellowship. 

Based on (1) observations obtained at the Gemini Observatory, which is 
operated by the Association of Universities for Research in Astronomy, Inc., under 
a cooperative agreement with the NSF on behalf of the Gemini partnership: the National 
Science Foundation (US), the Particle Physics and Astronomy Research Council (UK), the 
National Research Council (Canada), CONICYT (Chile), the Australian Research Council 
(Australia), CNPq (Brazil) and CONICET-Agencia Nac. de Promocion Cientifica y Tecnologica (Argentina). 
The Gemini Program ID is GN-2004B-Q-30; 

(2) data obtained at the W.M. Keck Observatory, which is operated as a
scientific partnership among the California Institute of Technology,
the University of California and the National Aeronautics and Space
Administration. The Observatory was made possible by the generous
financial support of the W.M. Keck Foundation;

(3) observations made through the Isaac Newton Groups' Wide Field Camera 
Survey Programme with the Isaac Newton Telescope operated on the island by the Isaac 
Newton Group in the Spanish Observatorio del Roque de los Muchachos of the Instituto de 
Astrofisica de Canarias.

The analysis pipeline used to reduce the DEIMOS data was developed at
UC Berkeley with support from NSF grant AST-0071048.

\clearpage
\begin{table}
\begin{center}
\caption{Observation log}
\begin{footnotesize}
\begin{tabular}{ccccrc}
\hline
\hline
Field  & $\alpha$  & $\delta$ & Date & Instrument &  Exposure time \\
& \multicolumn{2}{c}{(J2000)} & & & [s] \\
\hline
Field 1 & 10 09 01  & $+$12 22 47 & 2004 12 11 & GMOS &  3$\times$1800\\
	        	     &                  &                         & 2004 12 12 & GMOS & 3$\times$1800\\      
Field 2 & 10 08 59 & $+$12 16 00 & 2004 12 12 & GMOS & 3$\times$1800\\
	        	     &                  &                         & 2004 12 15 & GMOS & 3$\times$1800\\      
Field 3 & 10 08 39 & $+$12 19 43 & 2005 01 06 & GMOS & 2$\times$1800\\
\hline
\#1& 10 09 14 & $+$12 18 00 & 2005 02 06 & DEIMOS & 7200\\
\#2& 10 07 40 & $+$12 17 00 & 2005 02 07 & DEIMOS & 9000\\
\#3& 10 08 05 & $+$12 18 00 & 2005 02 08 & DEIMOS & 5400\\
\hline
\end{tabular}
\end{footnotesize}
\end{center}
\end{table}
\begin{table}
\begin{center}
\caption{Characteristics of the target stars. Note that the velocities from both the GMOS and DEIMOS samples 
have been shifted to a common mean.}
\begin{footnotesize}
\begin{tabular}{rccccccccccc}
\hline
\hline
{ID}   & {$\alpha$}  & {$\delta$} & {I} &  {V$-$I} & 
{$r$} & {$v_{HC}$} & {$\delta\,v_{HC}$} & 
{$R$} &  {S/N} &   {$[Fe/H]_{CG}$} &  {$\delta\,[Fe/H]_{CG}$} \\
       & \multicolumn{2}{c}{(J2000)} & & & ['] & \multicolumn{2}{c}{[km\,s$^{-1}$]} & & & \\
\hline
103 & 10 08 28  & 12 18 51 & 17.77 &  1.43 &  0.68 &  288.06 &  3.99 & 59.78 &  50 & $-$0.99 &  0.13\\
132 & 10 08 48  & 12 19 44 & 18.03 &  1.48 &  5.15 &  282.07 &  6.28 & 36.12 &  28 & $-$1.45 &  0.11\\
137 & 10 08 27  & 12 20 31 & 17.93 &  1.56 &  2.80 &  292.81 &  5.33 & 43.99 &  40 & $-$1.38 &  0.12\\
141 & 10 09 00  & 12 18 21 & 18.09 &  1.43 &  7.91 &  276.51 &  3.97 & 58.39 &  65 & $-$1.76 &  0.11\\
143 & 10 08 32  & 12 18 34 & 18.10 &  1.32 &  1.18 &  292.99 & 14.29 & 14.10 &   6 & \nodata &   \nodata\\
\hline
\end{tabular}
\end{footnotesize}
\end{center}
\begin{footnotesize}
{Note. --- This Table is published in its entirety in the electronic edition of the {\it 
Astrophysical Journal}. 
A portion is shown here for guidance regarding its form and content.}
\end{footnotesize}
\end{table}
\begin{table}
\begin{center}
\caption{Mean velocities and dispersions for different fields.}
\begin{footnotesize}
\begin{tabular}{lcr}
\hline
\hline
{}   & {$\left<v\right>$} & {$\sigma$} \\
{\raisebox{1.5ex}[-1.5ex]{Field}}   & {[\,km\,s$^{-1}$]} & {[\,km\,s$^{-1}$]} \\
\hline
GMOS,  Field 1 & 284.0 $\pm$ 2.6 & 7.4 $\pm$ 3.4 \\
GMOS,  Field 2 & 283.6 $\pm$ 2.8 & 8.8 $\pm$ 3.8 \\
GMOS,  Field 3 & 284.2 $\pm$ 2.3 & 12.5 $\pm$ 3.6 \\
GMOS, all & 284.6 $\pm$ 1.5 &9.9 $\pm$ 2.1 \\
DEIMOS, all &  283.8 $\pm$ 1.5 & 9.9 $\pm$ 2.1 \\
GMOS $+$ DEIMOS &  284.2 $\pm$ 1.0 & 9.9 $\pm$ 1.5 \\
GMOS $+$ DEIMOS $+$ M98 & 287.0 $\pm$ 0.9 & 9.8 $\pm$ 1.3 \\
\hline
\end{tabular}
\end{footnotesize}
\end{center}
\end{table}
\begin{table}
\begin{center}
\caption{Mass and central density estimates.}
\begin{footnotesize}
\begin{tabular}{llccl}
\hline
\hline
{}   & {} &  {$M [r<r_{tid}]$} & {$\rho_0$} & {($M/L_V)$} \\
 {\raisebox{1.5ex}[-1.5ex]{Dispersion profile}} &  {\raisebox{1.5ex}[-1.5ex]{Sample}}  
& {[10$^7\,M_{\odot}$]} & {[$10^8\,M_{\odot}\,kpc^{-3}$]} &  {[($M/L_V)_{\odot}$]}  \\
\hline
Flat &  & 8.7$^{+1.6}_{-2.6}$ & 2.7$^{+0.6}_{-0.9}$ & 26$^{+5}_{-8}$\\
Falling &GMOS + DEIMOS& 8.0$^{+2.7}_{-2.4}$ & 2.8$^{+0.9}_{-0.9}$ & 24$^{+7}_{-6}$\\
Peaked &  & 7.7$^{+2.4}_{-1.6}$ & 2.9$^{+0.9}_{-0.6}$ & 23$^{+7}_{-5}$\\
\hline
Flat &  & 7.9$^{+2.4}_{-1.0}$ & 2.8$^{+0.8}_{-0.4}$ & 23$^{+7}_{-3}$\\
Falling &GMOS + DEIMOS + M98& 7.8$^{+2.4}_{-1.0}$ & 2.8$^{+0.8}_{-0.4}$ & 23$^{+7}_{-3}$\\
Peaked &  & 4.8$^{+2.7}_{-1.8}$ & 2.0$^{+1.2}_{-0.8}$ & 14$^{+8}_{-6}$ \\
\hline
\end{tabular}
\end{footnotesize}
\end{center}
\end{table}
\clearpage
\begin{figure}
\plotone{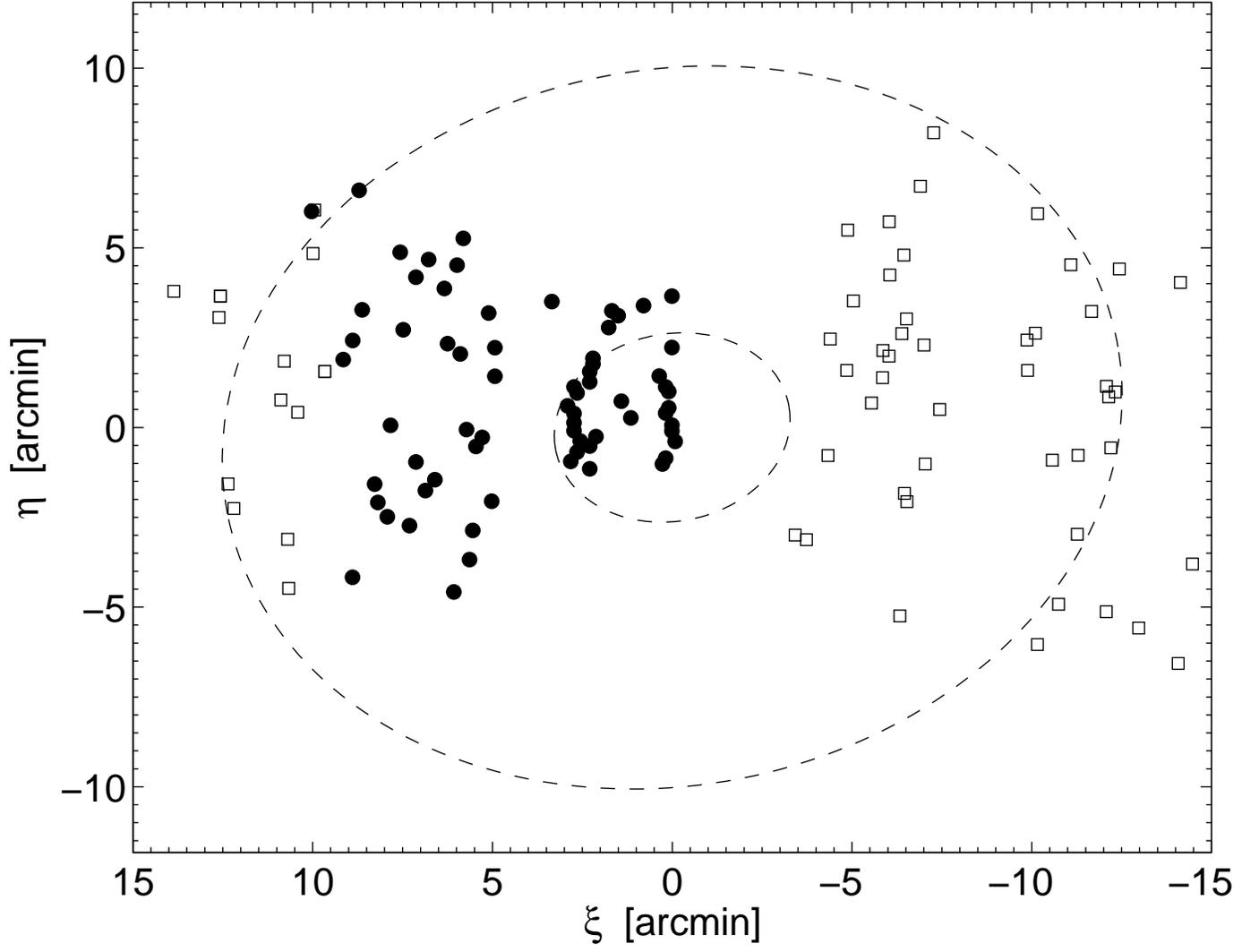}
\caption{Location of our targets centered on the LeoI dSph at
($\alpha,\delta)=($10:08:28, +12:18:18).  Red giants observed with
GMOS are shown as filled circles, DEIMOS targets are depicted as open
squares. Overlaid ellipses designate respectively Leo\,I's core and
nominal tidal radius at 3$\farcm$3, 12$\farcm$6, respectively. 
The system's ellipticity and position angle are assumed to be
0.21 and 79$\degr$.} 
\end{figure}
\begin{figure}
\plotone{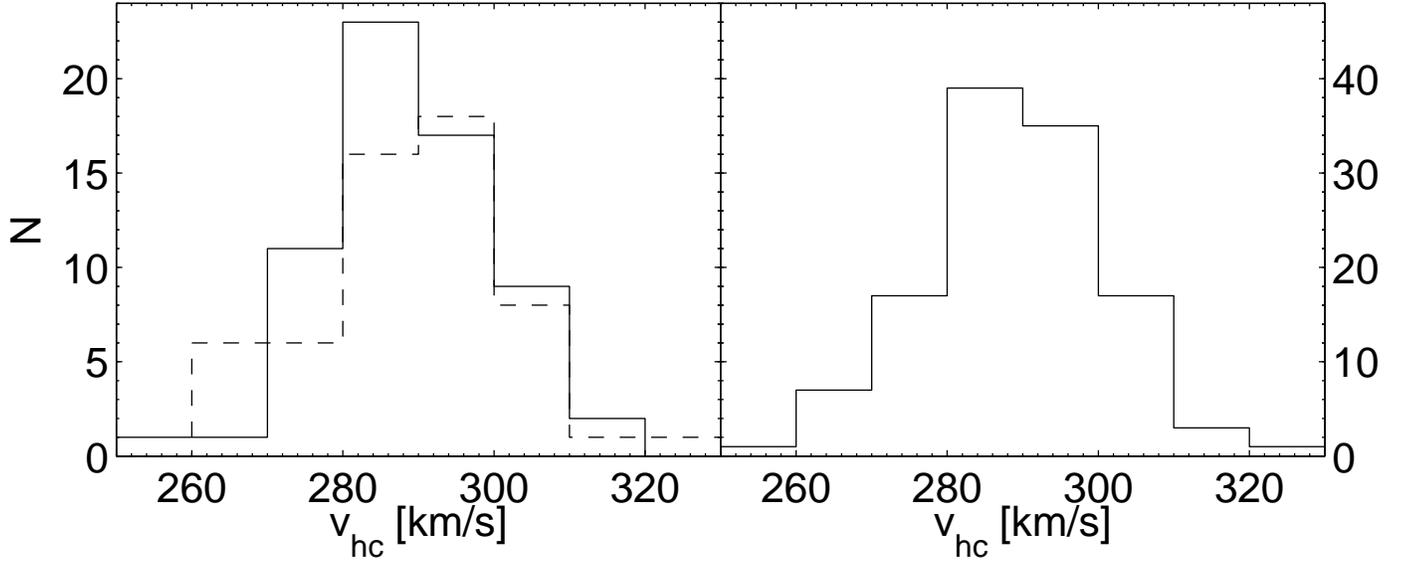}
\caption{Radial velocity histograms of the GMOS data (solid line in
the left panel), DEIMOS stars (dashed line, left panel) and combined
GMOS plus DEIMOS set (right panel). All targets with velocities larger
than 200\,km\,s$^{-1}$ are shown. Within this range, two targets (at
251 and 324 km\,s$^{-1}$, respectively) fall outside a
$\pm3\sigma$-cut. }
\end{figure}
\begin{figure}
\plotone{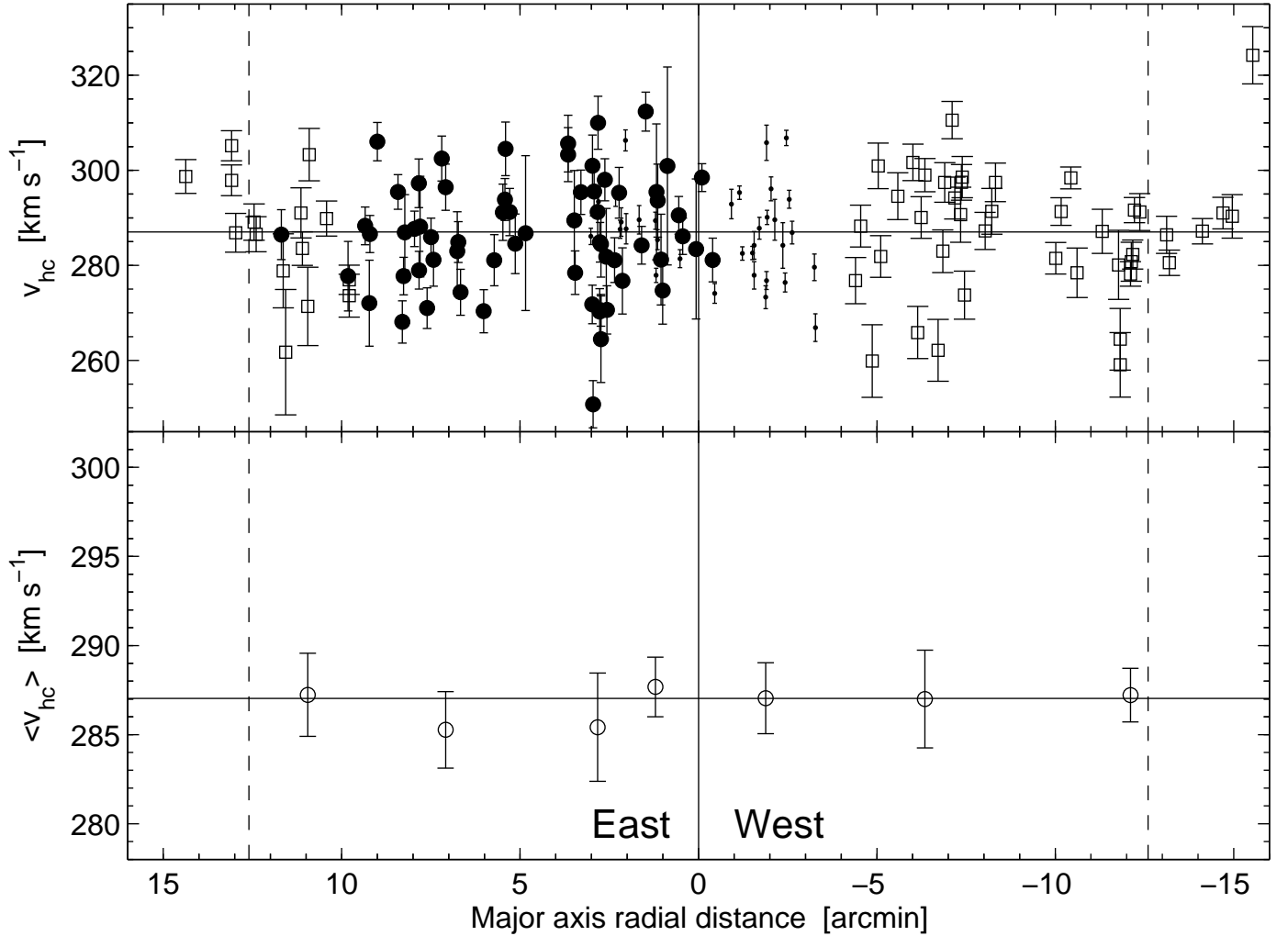}
\caption{Distributions of individual radial velocities (top panel) and 
mean radial velocities (bottom panels) as a function of their major axis distance. 
The symbols in the upper panel are the same as in Fig.~1, while the small dots additionally 
depict the 33 data points from M98. The solid horizontal line indicates the global mean of our 
sample, while the dashed lines denote the formal King radius at 12$\farcm$6}
\end{figure}
\begin{figure}
\plotone{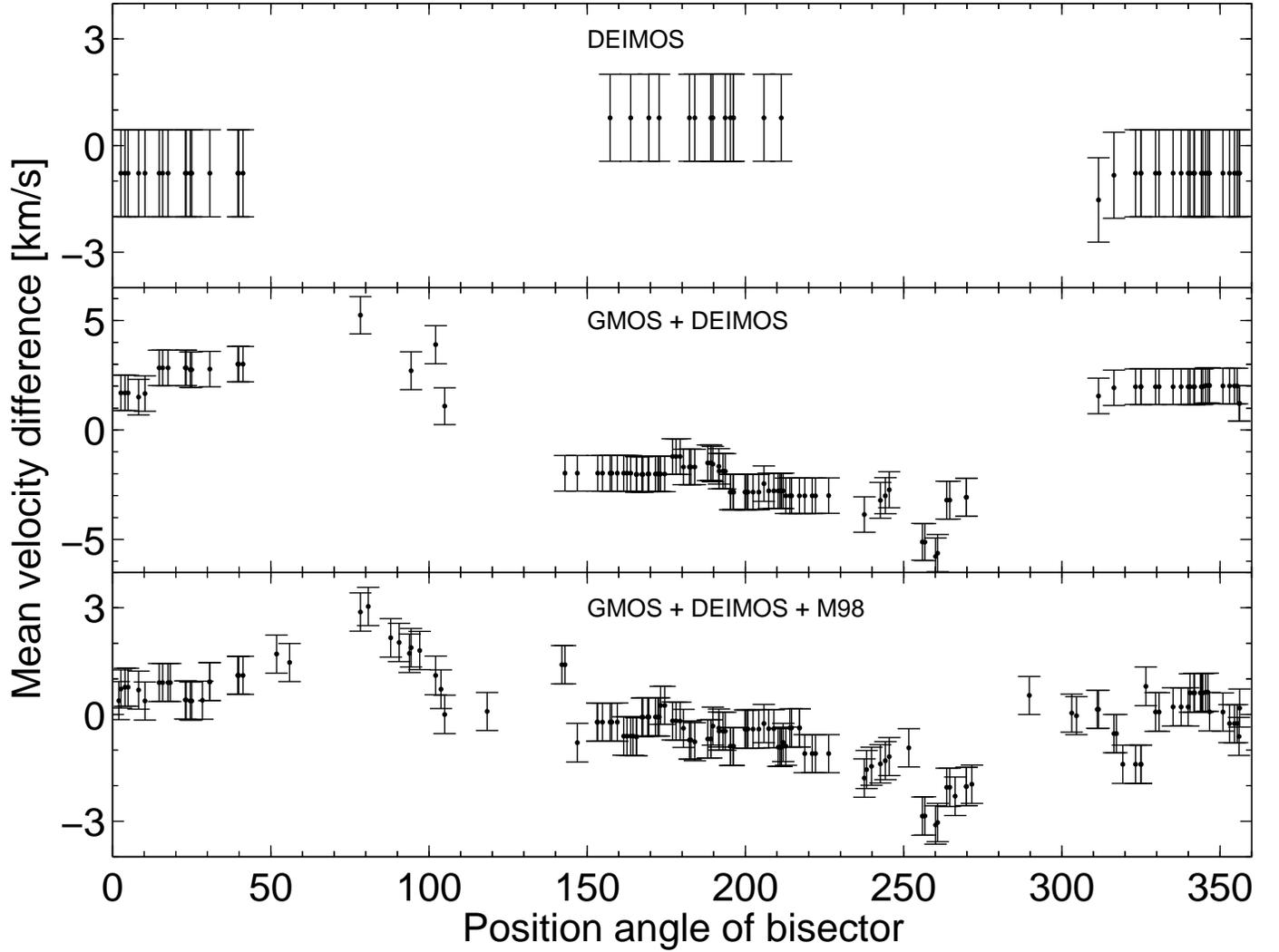}
\caption{Tests for rotation in different subsamples of our data (as
labelled): shown is the mean velocity difference with respect to
rotation around axes at the respective position angles.  A rotation
signal around the major axis (at PA$\sim 79\degr$) is distinguishable,
but its significance is marginal. See text for a detailed discussion.}
\end{figure}
\begin{figure}
\plotone{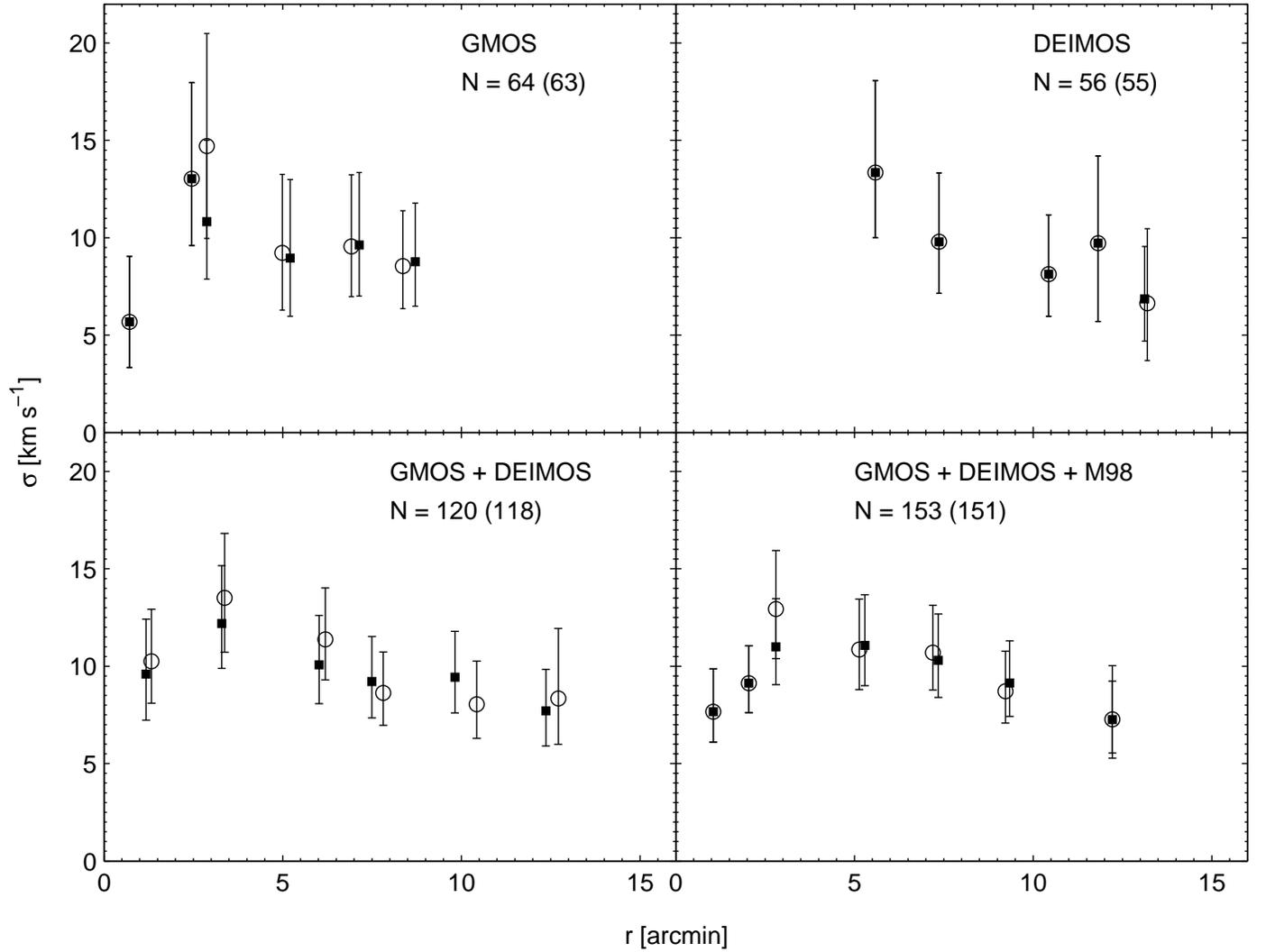}
\caption{Velocity dispersion profile for different subsamples, as
labelled, and rejection cuts.  The numbers given refer to the sample
size.  Open circles refer to the full samples, whereas filled squares
as well as the numbers in parentheses relate to the data cut at
3$\sigma$.}
\end{figure}
\begin{figure}
\plotone{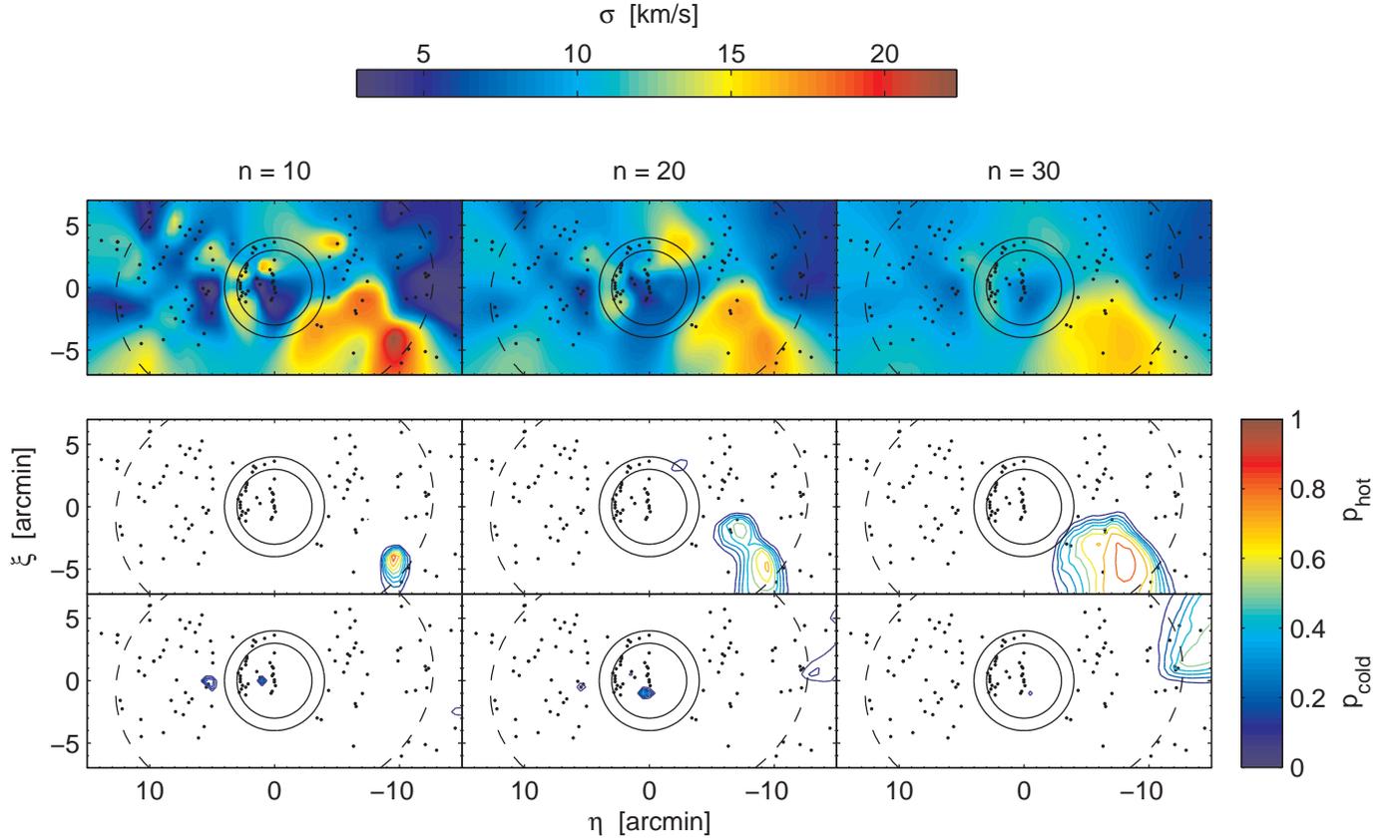}
\caption{Top panel: Radial velocity dispersion estimator according to
eq. (3), following the method of Walker et al. (2006b).  Shown are the
data for the combined GMOS + DEIMOS samples (small dots are the target
locations), but the results do not alter after inclusion of the M98
sample. The middle and bottom panels display contours of the
statistical significances $p_{hot}$ and $p_{cold}$ for the occurrence
of any kinematically hot (middle) or cold (bottom) substructure. The
individual panels show results for different choices of neighboring
points included in the dispersion estimates.  Contours are shown in
intervals of 0.1.  The dashed ellipse circumfers the nominal tidal
radius and the two solid lines inscribe the interval of
3$\arcmin$--$4\arcmin$, in which the radial dispersion profile
exhibits a bump.  See text for details. [{\em See the electronic edition 
of the Jounal for a color version of this figure.}]}
\end{figure}
\begin{figure}
\plotone{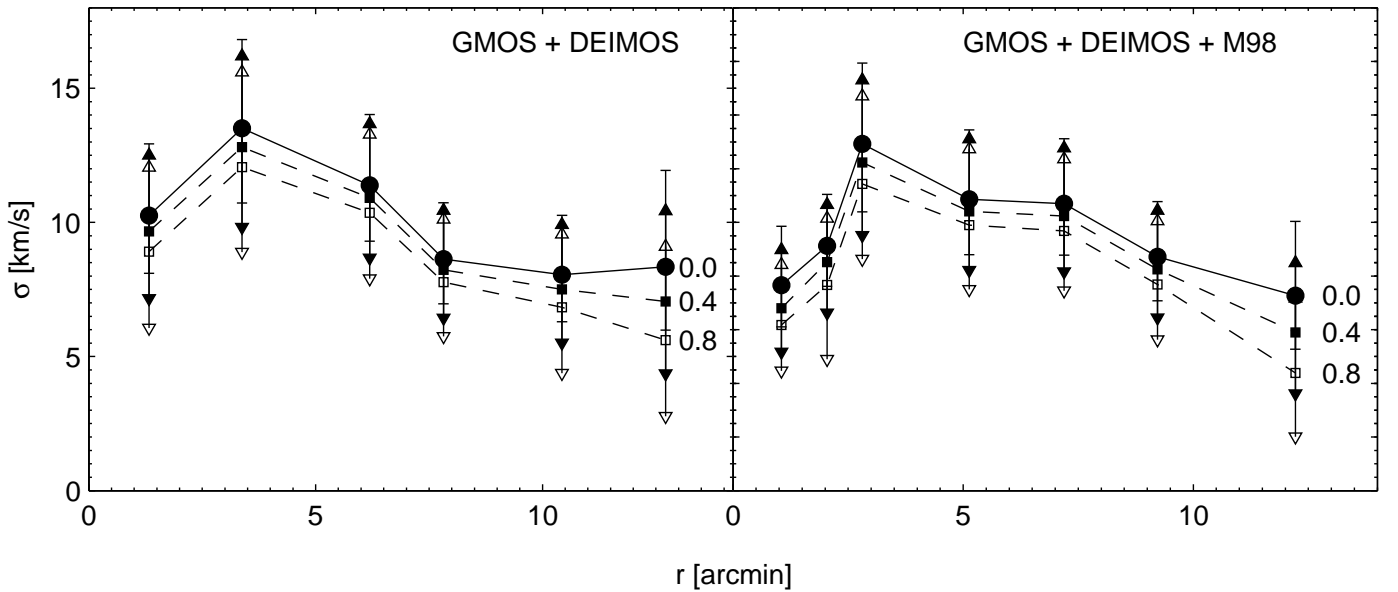}
\caption{Radial velocity dispersion profiles for the subsamples as in
Fig.~5. From top to bottom each curve was calculated with increased
binary fractions $f_b$ (labelled right of the outer bins). The solid
curve and filled points refer to the observed profile assuming no
binaries. Solid (open) squares and dashed lines are drawn from a
maximization assuming an $f_b$ of 0.4 and 0.8, respectively.  In order
to illustrate the effect of a binary population on the errors on the
velocity dispersion, the respective error bounds of the profiles with
$f_b>0$ are displayed as solid (open) triangles.}
\end{figure}
\begin{figure}
\plotone{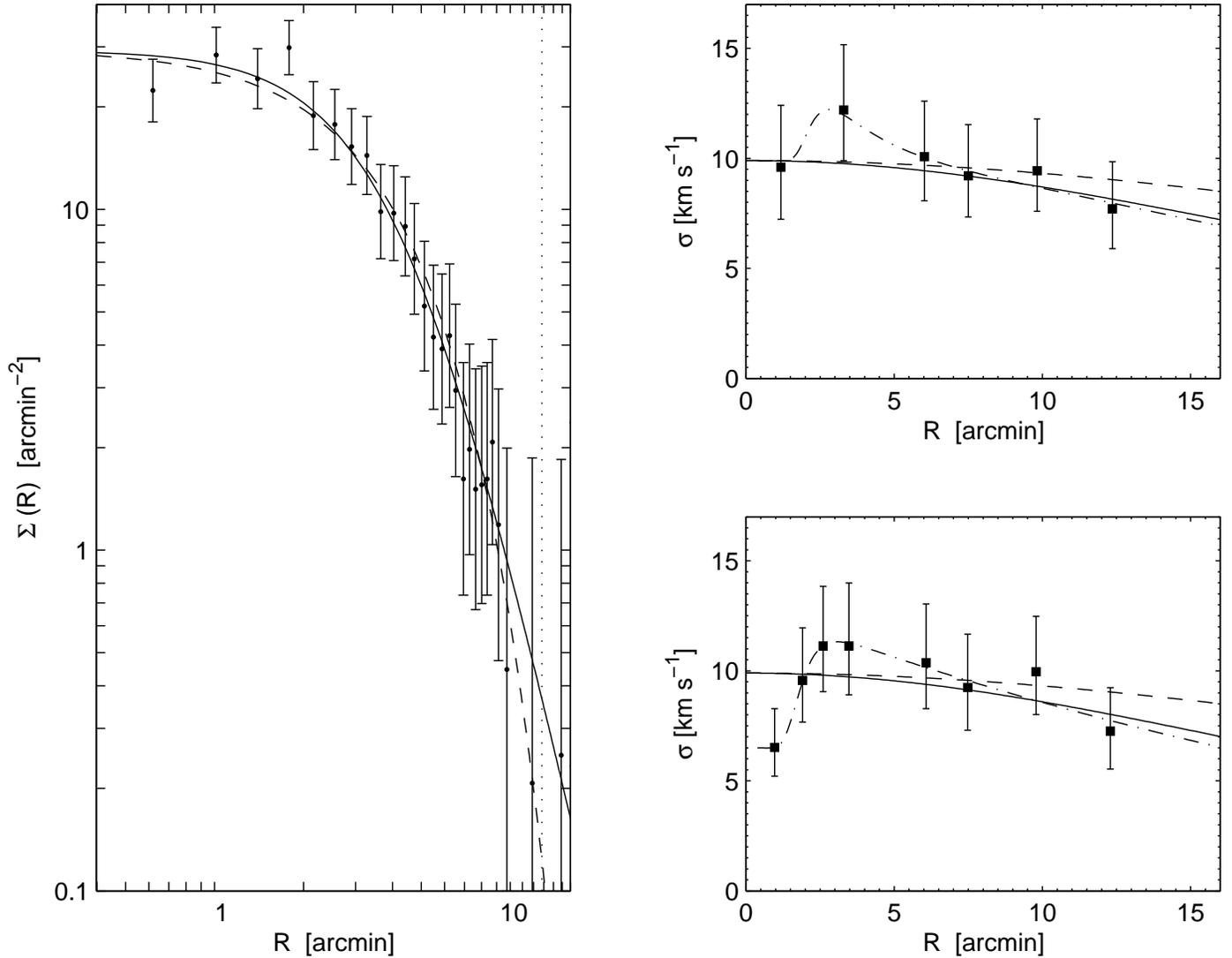}
\caption{Left panel: Surface brightness profile of Leo\,I from Irwin
\& Hatzidimitriou (1995) together with the fit of a Plummer law (solid
line) and an exponential fit (dashed). The upper right panel shows
fits to the dispersion profiles from the GMOS plus DEIMOS sample, and
the analog after inclusion of M98's data is displayed in the lower
right panel. Overplotted are fits of Plummer laws with different scale
radii such as to reproduce a close-to flat (solid line) or falling (dashed)
profile and also a modified profile, which incorporates the bump at $3\arcmin$ (dash-dotted).}
\end{figure}
\begin{figure}
\plotone{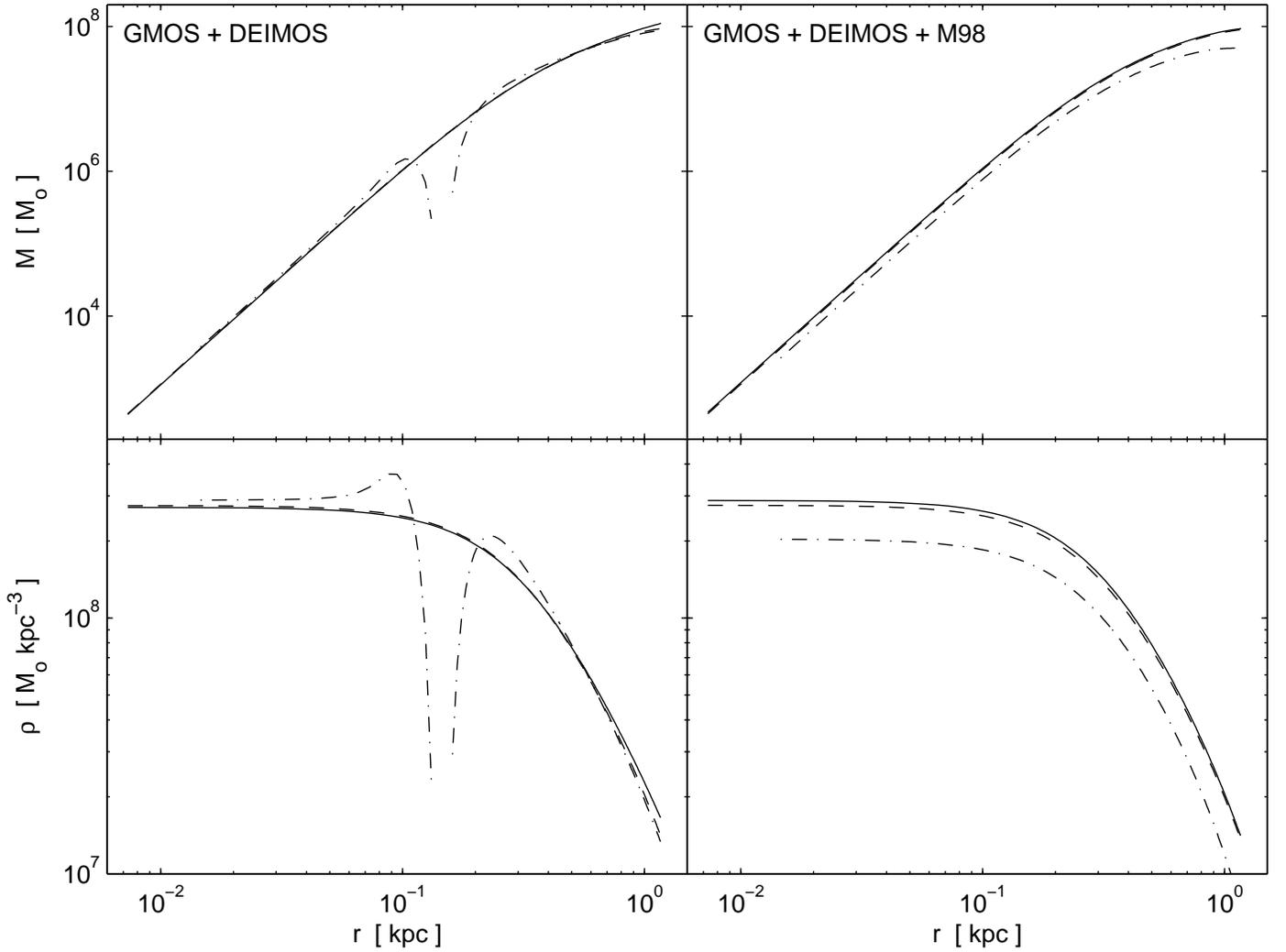}
\caption{Mass (upper panel) and density (lower panel) estimates from
Jeans equation, assuming an isotropic velocity distribution. The left
panels each refer to the GMOS $+$ DEIMOS sample, whereas the right
panels additionally account for the M98 data set. Different line types
signify different approaches to fit the observed dispersion profile
(see Fig.~8).  The unphysical drop in the profiles of the left panels
is due to the inclusion of an additional peak in the best-fit velocity
dispersion profile, which exhibits a steep rising gradient in the
inner regions opposed to a slower decline of the underlying Plummer
profiles. These counteracting gradients are not as distinct after
inclusion of the M98 sample so that the right panels do not display
this drop.}
\end{figure}
\begin{figure}
\plotone{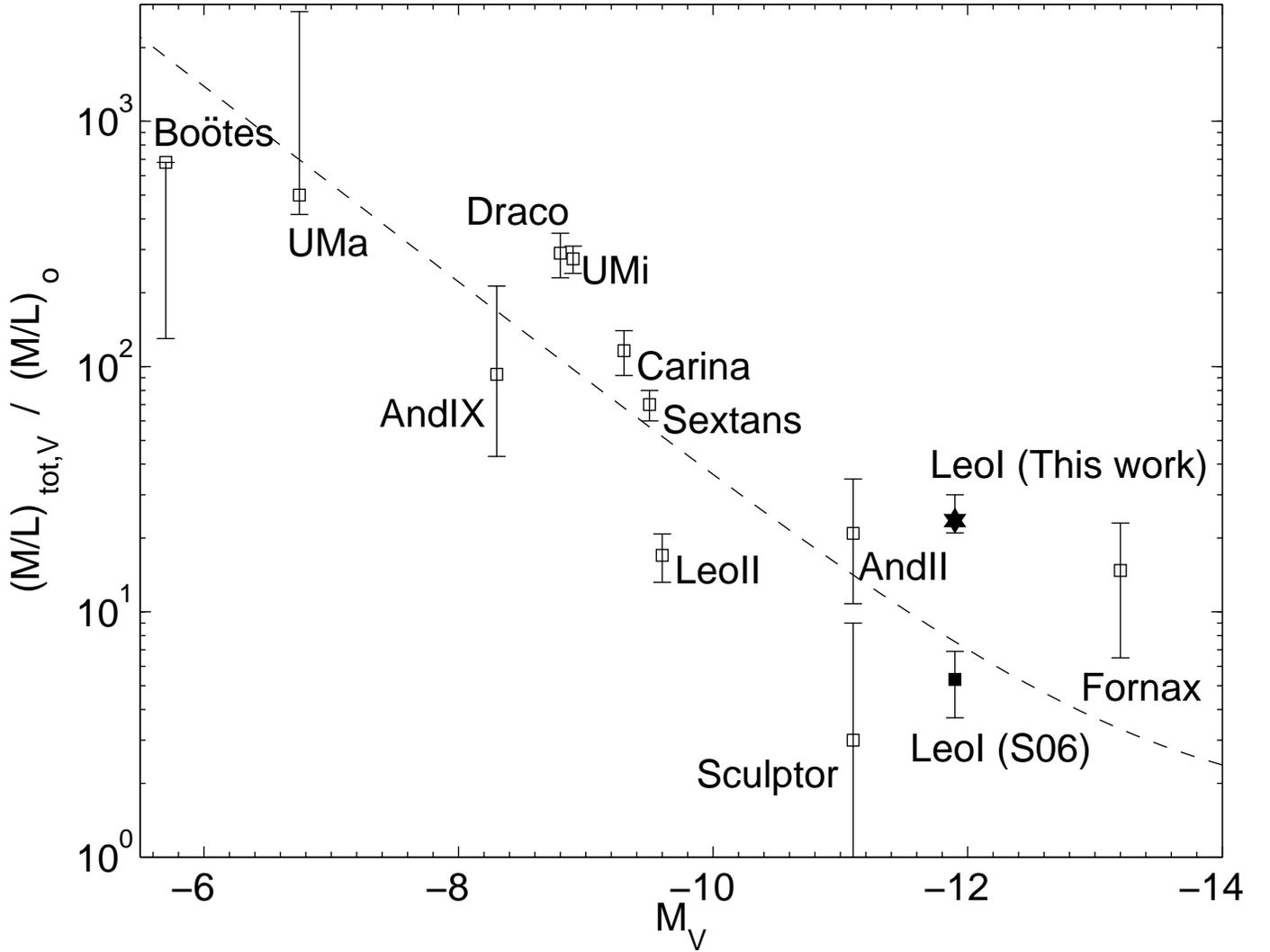}
\caption{Mass-to-light ratios for LG dSphs (after Wilkinson et
al. 2006a).  The most recent individual mass estimates are from the
sources cited in the text.  The dashed line assumes a stellar
(M/L)$_V$ of 1.5 in solar units and a dark matter component of
3$\times 10^7$M$_{\odot}$. Also shown for Leo\,I is the recent estimate 
of S06, who yield a lower value than the present work.}
\end{figure}
\begin{figure}
\epsscale{.8}
\plotone{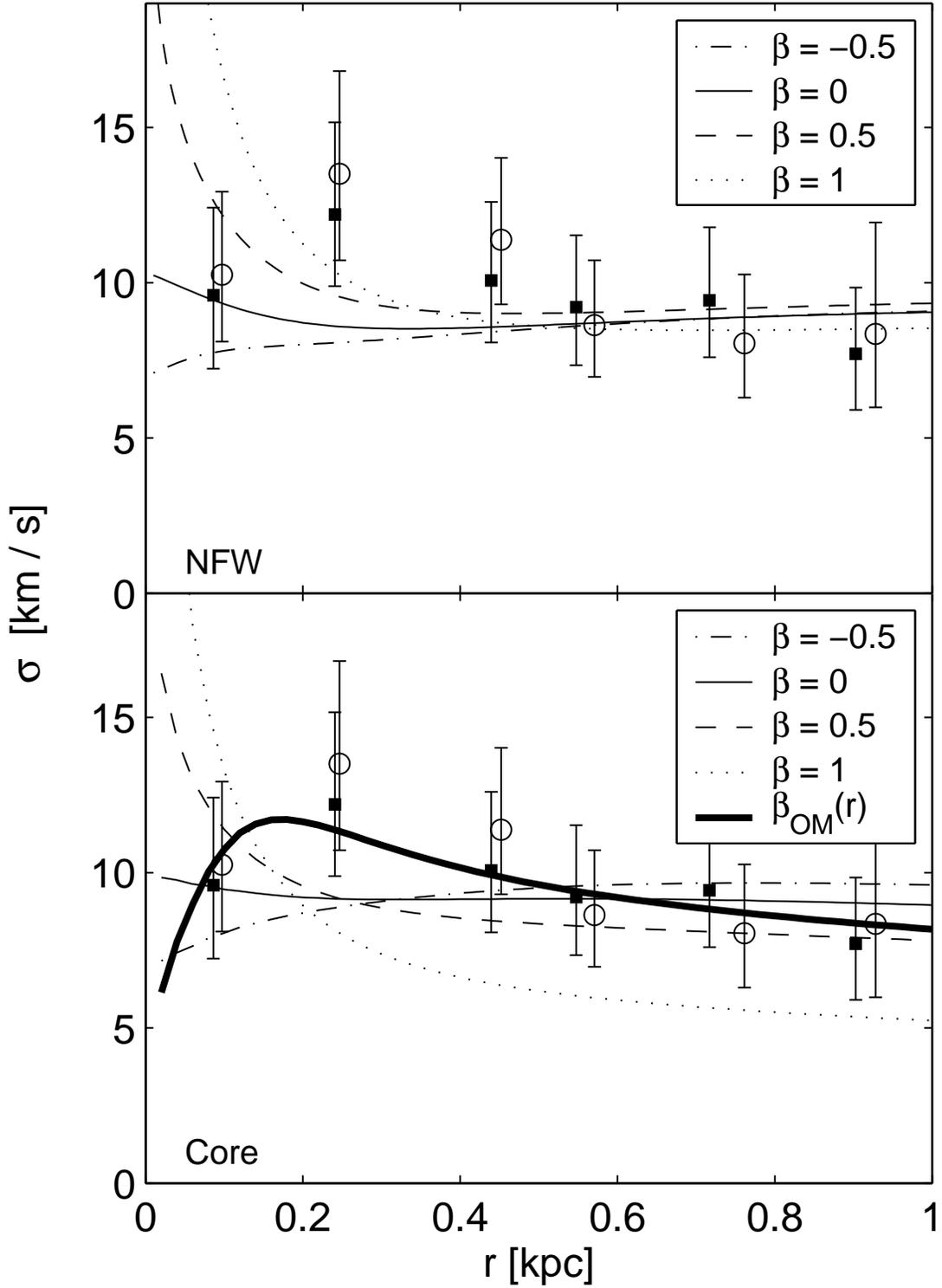}
\epsscale{1}
\caption{Predicted velocity dispersion profiles with varying degree of
anisotropy ($\beta$) for two different halo density profiles, namely,
cuspy (``NFW'', top panel) and a cored halo (bottom panel).  The inner
regions are consistent with isotropic orbits, whereas the outer
regions do not rule out a certain amount of radial anisotropy. The lower panel also 
contains a curve for the case of a radial varying anisotropy parameter, $\beta_{\rm OM}$, 
after Osipkov (1979) and Merrit (1985), which represents our data reasonably well. 
}
\end{figure}
\begin{figure}
\plotone{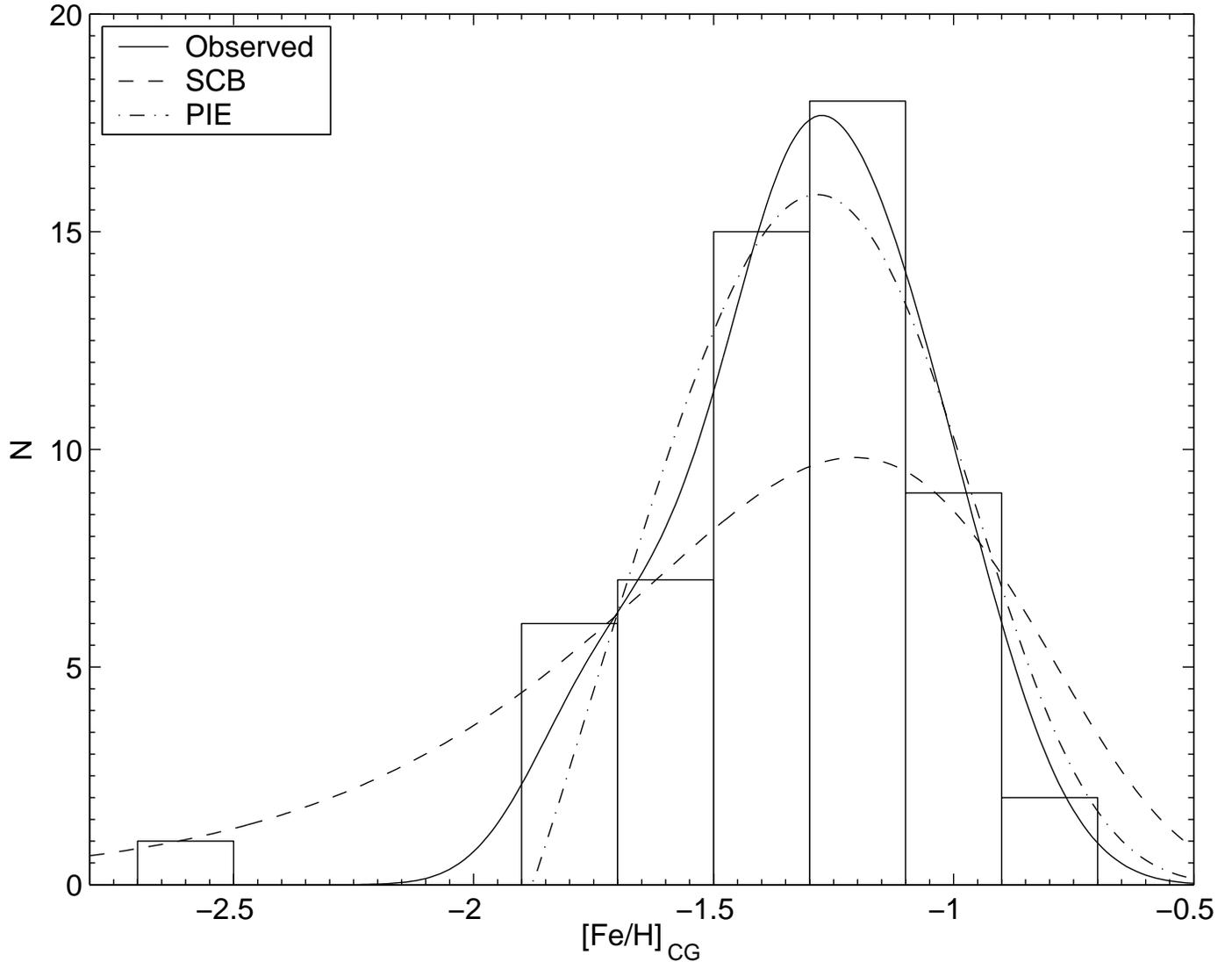}
\caption{Metallicity distribution for the 58 GMOS radial velocity
member stars, for which EWs could be measured. These were convolved by
observational uncertainties to yield the solid curve.  Overplotted are
a modified simple closed-box model (SCB, dashed line) and a model
using Prompt Initial Enrichment (PIE, dash-dotted line), scaled to the
same number of stars. }
\end{figure}
\begin{figure}
\plotone{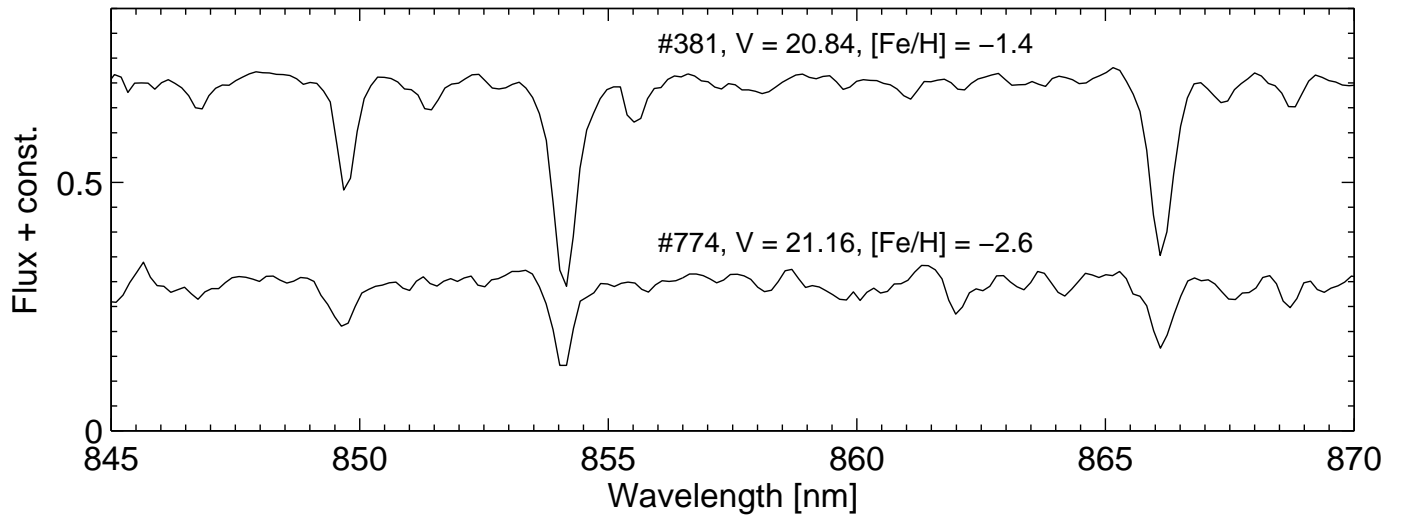}
\caption{Spectra of the most metal poor star in our sample and a
typical spectrum of a target around the peak metallicity. Apart from
the three dominant Ca lines there are a few weak neutral metal lines
detectable, which are not distinguishable from the continuum in the
metal poorer target.}
\end{figure}
\begin{figure}
\plotone{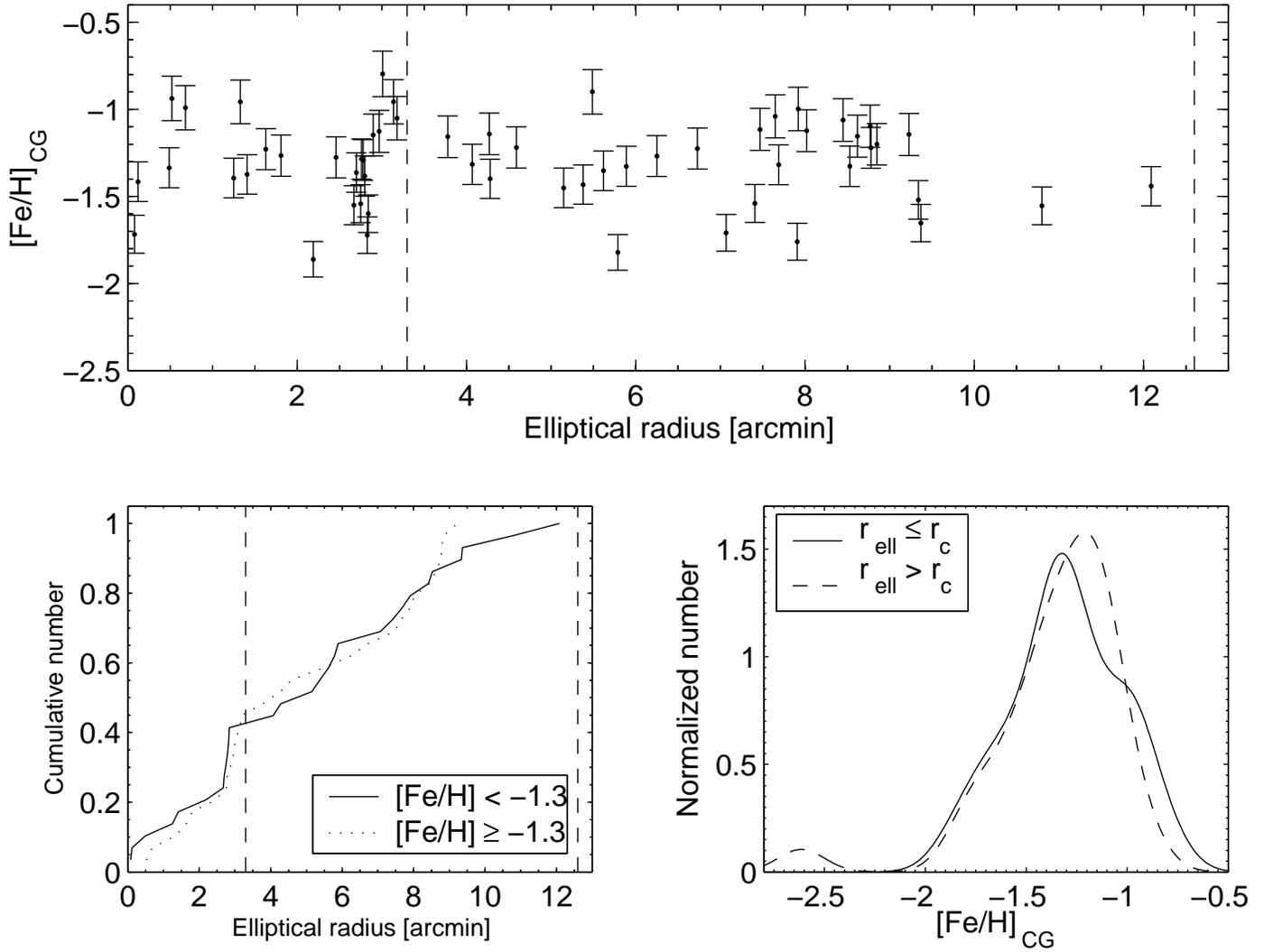}
\caption{CaT metallicities of our radial velocity members vs. their
elliptical radius (top). Nominal core and tidal radius are also
denoted by vertical  lines. The bottom left shows the cumulative
number distribution versus radius of the metal richer component (solid
line) and the metal poorer population (dotted line).  These are
practically indistinguishable.  MDFs at different radii are displayed
in the bottom right panel. These density distributions were convolved
by individual measurement errors. No apparent trend of [Fe/H] with
radius is discerned.}
\end{figure}

\end{document}